%% file: opstal.tex
\documentstyle[12pt, epsf]{article}

\def \s{\sigma}

\def \spin{{\rm spin}}
\def \matrix{{\rm matrix}}
\def \m#1{$#1$}
\def\hatleftix{{\hat{\mathit L}_{\Lambda}}}
\def\hatrightix{{\hat{\mathit R}_{\Lambda}}}
\def\leftix{{{\mathit L}_{\Lambda}}}
\def\rightix{{{\mathit R}_{\Lambda}}}
\def\hatcentrix{{\hat{\mathit \Sigma}_{\Lambda}}}
\def\centrix{{{\mathit \Sigma}_{\Lambda}}}
\def\vectrix{{\cal V}_{\Lambda}}
\def\grandhatL{{\hat{\mathit L}_{\Lambda,\Lambda_F}}}
\def\hatmultix{{\hat{\mathit M}_{\Lambda}}}
\def\multix{{{\mathit M}_{\Lambda}}}
\def\salt{{\mathit F}_{\Lambda}}
\def\saltone{{{\mathit F}_1'}}
\def\grandsalt{{\hat{\mathit F}_{\Lambda,\Lambda_F}}}
\def\ignore#1{}
\def \gl{gl_{+\infty}}

\newcommand\beq{\begin{eqnarray}}
\newcommand\eeq{\end{eqnarray}}
\newcommand{\beqs}{\begin{eqnarray}}
\newcommand{\eeqs}{\end{eqnarray}}

\begin{document}
\begin{center}
   {\LARGE\bf  Symmetry Algebras of Large-$N$ Matrix Models for Open Strings} \\
   \vspace{2cm}
   {\large\bf C.-W. H. Lee and  S. G. Rajeev} \\
   {\it Department of Physics and Astronomy, University of Rochester, 
    Rochester, New York 14627} \\
   \vspace{.5cm}
   {December 9th, 1997} \\
   \vspace{2cm}
   {\large\bf Abstract}
\end{center}

We have discovered  that the  gauge invariant observables of matrix models
invariant under U($N$) form a Lie algebra, in the  
planar large-$N$ limit. These models include 
 Quantum Chromodynamics and the  M(atrix)-Theory of strings. We study here
the gauge invariant states corresponding to  {\it open}  strings
 (`mesons').  We find that the algebra is an extension of a remarkable new Lie
algebra  \m{{\cal V}_\Lambda} by a product of more well-known algebras
such as  \m{\gl} and the Cuntz algebra.  \m{{\cal V}_\Lambda} appears to be a 
generalization of the Lie algebra of
vector fields on the circle to non-commutative geometry.  We also
use a representation of our Lie algebra to 
establish an isomorphism between certain matrix models (those that
preserve `gluon number')  and open quantum  spin chains. Using known
results from quantum spin chains,
we are able to identify some exactly solvable matrix models. Finally, the
Hamiltonian of a dimensionally reduced  QCD model is expressed explicitly as an
element of our Lie algebra.

\begin{flushleft}
{\it PACS}: 11.25.Hf, 11.15.Pg, 02.20.Sv, 75.10.Jm.\\
{\it Keywords}: open strings, matrix models, gauge theories, 
current algebra, quantum spin chains, large-$N$ limit.
\end{flushleft}
\pagebreak

\section{Introduction}

One of the major problems in physics  is to understand Yang-Mills 
theories such as  Quantum ChromoDynamics (QCD). It is natural to look for
a classical approximation to this difficult problem.  The obvious  classical limit is to 
let Planck's constant $\hbar$ go to 0.  Unfortunately, this limit is 
ill-suited for describing long-distance phenomena such as  quark confinement.
Another, less obvious,classical limit   is to let the number of colors $N$ go to 
$\infty$.  This large-$N$ limit smooths out quantum fluctuations of
gauge-invariant observables, and 
is believed to manifest not only asymptotic freedom but also quark confinement
and so should describe long-distance interactions among quarks better.

We obtained the Poisson algebra of observables in this large-$N$ limit in a 
previous paper \cite{rajeevturgut}. It was highly nonlinear, reflecting the 
dynamical fact that the large-$N$ limit of QCD is  a nonlinear classical 
theory.  A further approximation is needed.
This is the theory of small oscillations around the ground  state, which can 
also be understood as the sum of planar diagrams in perturbation theory. In this 
paper we will study this `planar large-$N$ limit' of matrix models
such as QCD.  We will not use directly the results of Ref.\cite{rajeevturgut}; 
instead we will give a direct argument
starting with the canonical (anti)-commutation relations.

In a previous Letter \cite{leerajlett}, we reported on a Lie algebra describing the
symmetry of closed string states like glueballs.  The present paper serves as a mathematical
foundation for the results in that Letter.  Here we will deal exclusively with open string
states.  Building upon the mathematical results here, we will turn to closed string states, the mathematics
of which turns out to be more tricky, and give a detailed 
exposition of the results in that Letter in the next paper \cite{clstal}.

We will not study the problems of Yang-Mills theories directly here;
instead we will look at some matrix models for Yang-Mills
theories. In this sense the present work fits into the program of
studying Universal Yang-Mills Theories \cite{uym}. The case
with fermionic matter fields in Universal Yang-Mills theory was
studied in Ref.\cite{rajfer}. In our
present language Universal Yang-Mills Theory corresponds to the case
\m{\Lambda=1} with just one matrix degree of freedom. This 
thus generalizes the considerations of Ref.\cite{rajfer} to the
case where there are several matrices modeling gluons. For an earlier
attempt to consider multi-matrix models from this point of view, see Ref.\cite{rajfergeom}.

It is well known \cite{thooft, wittenN} that in the planar limit mesons and glueballs 
are stable, non-interacting particles. Gauge-invariant observables, such as the Hamiltonian, act as linear 
operators on these states in the light-cone formalism \cite{bpp} in which (unlike in some other
quantization schemes) the ground state is the same as that in free theory --- we do not need to worry about 
non-perturbative zero mode contributions. We find that the gauge-invariant observables form a Lie algebra in this 
limit. Just as the representation theories of Kac-Moody and Virasoro algebras allow us
to solve conformal field theories, we believe that the study of our
Lie algebra  will yield a solution to  Yang-Mills theory in the planar limit.
Indeed any attempt to understand  the spectrum of QCD analytically 
has to begin with a solution of the planar limit. This means that
we need to identify the symmetries peculiar to this limit and learn to
exploit them to the fullest. 

The idea that a dynamical symmetry will help solve for the spectrum of
a quantum theory is familiar from other problems of theoretical
physics. Perhaps the oldest examples are the hydrogen atom and the
harmonic oscillator. In the former, the dynamical symmetry is
SO(4) and the hamiltonian can be written as a rational  function
of  the generators of this Lie algebra. The representation theory of
SO(4) then yields the spectrum. (This was Pauli's solution of the
hydrogen atom, done even before the Schrodinger equation was
discovered.) In the case of the harmonic socillator the symmetry is
the symplectic Lie algebra sp(2), which is the algebra of
derivations of the canonical commutation relations. The Hamiltonian is
an element of sp(2). The familiar solution of the harmonic
oscillator in terms of creation-annihilation operators has exactly
this meaning. If we understand the dynamical symmetry of planar QCD equally
well, we would be able to solve for its spectrum by a similar method.
We will give an example of this in a simplified model of QCD, in a
variational approximation.

Conformal field theory  provides more recent examples of the solution of a
quantum theory by Lie algebraic methods \cite{wzw}.
 The underlying  symmetry of the Wess-Zumino-Witten model (for
example) is the semi-direct product 
of  the Kac-Moody algebra by the Virasoro algebra; the latter acts
as derivations on the former. The Hamiltonian is an element of the
Virasoro algebra. The partition
function of the WZW model can then be obatined from the Kac-Weyl
character formula: this determines the eignvalues of the Hamiltonian. 
The corresponding wavefunctions can be understood as sections of the
determinant line on coset spaces of the Loop Group \cite{prseg}.

We are at the moment a long way from such a complete understanding of
planar QCD. In this paper we take a crucial step in this direction: we
identify its  symmetry algebra. Somewhat surprisingly we find that it
has a structure reminiscent of conformal field theory. Indeed it turns
out that planar QCD is a
non-commutative generalization of conformal field theory. The analogue
of Kac-Moody algebra is the Cuntz algebra \cite{3, 4}: it is the current
algebra of planar QCD\footnote{Previously, Gopakumar and Gross found that the Cuntz algebra was relevant to the 
master field \cite{5}.  Even in this special case, our discussion will be more complete and direct, being in 
the Hamiltonian rather than the perturbative formalism.  Another previous work on
relating the Cuntz algebra and the large-$N$ limit was done by Turgut \cite{turgut}.}.

There is even a natural chiral structure for the current algebra we find.
There are two commuting (up to extensions) copies of the Cuntz algebra
that act on the left end (the end containing an antiquark) 
and right end (the one with a quark) of a meson state. We discover the
analogue of the Virasoro algebra \cite{7} which we will call the centrix
algebra. It consists of operators that act on the gluons in the meson
state. It acts as derivations on the Cuntz  algebra, exatly like the
action of the Virasoro algebra on the Kac-Moody algebra. Indeed, in  the
special case of one degree of freedom for the gluons, our algebras
just reduce to the Kac-Moody-Virasoro algebra. Even this special
case is useful: it determines the spectrum of some models for QCD
within a variational approximation. 

We discover that  the symmetry algebra of planar
QCD has ideals isomorphic to the algebra of finite rank
matrices. This shows that the {\em essential spectrum} of the planar QCD
hamiltonian is determined by the quotient algebra. Thus the essential
spectrum may be easier to determine; this is also suggested by the
analogy with the theory of Toeplitz operators. Thus our mathematical
study suggests some strategies for the kind of physical approximations
we should make in solving planar QCD.

It has been suspected at least since 't Hooft's paper \cite{thooft} that planar
Yang-Mills theory is closely related to a string theory. Indeed, recent works
\cite{ferretti, susskind} indicates that M-theory, the latest incarnation  of
string theory, is a Matrix model. Thus our Lie algebra is also the symmetry 
algebra of M-theory. A special case of our Lie algebra is just the Virasoro 
algebra, suggesting that our  theory is indeed a generalization of string 
theory. 

We will find that the Lie algebra of observables has several ideals
isomorphic to well-known algebras such as \m{\gl} and Cuntz
algebras and their products. After we quotient by these ideals we will find that what is
left over is a remarkable new Lie algebra \m{{\cal V}_{\Lambda}}. When
\m{\Lambda=1} this is just the Lie algebra of (polynomial ) vector
fields on a circle, the Virasoro algebra, up to an extension.  In the
general case it is,  roughly speaking,  the Lie algebra of outer
deruvations of the Cuntz algebra.  Now, 
the Cuntz algebra is a non-commutative generalization of the algebra of
functions on a circle; and derivations are non-commutative
generalizations of vector fields. 
Thus we can think of \m{{\cal V}_{\Lambda}} as the algebra of vector 
fields on a non-commutative manifold, whose algebra of functions is Cuntz 
algebra. This interpretation is particularly striking in the context
of $M$-theory and its connection to quantum theories of  gravity. 

If we have found an essential new symmetry algebra, it ought to help
us solve exactly some matrix models. We are in fact able to do this by
using our  Lie algebra to prove an isomorphism between certain classes of
matrix models (those that preserve `gluon number') and quantum spin chain systems. By
translating known exactly solvable spin chains into the language of
matrix  model
we identify some solvable matrix models. In the case of open
spin chains these are solutions to the Yang-Baxter and Sklyanin
relations \cite{baxter, Sk, GoRuSi}. 

The synopsis of this paper is as follows.  In Section~\ref{s2}, we will give a formulation of the physical 
states and the operators in the large-$N$ limit.   Then, in 
Section~\ref{s3}, we will decsribe the current algebra in the large-$N$
limit.  It has two parts --- the leftix and rightix algebras of operators ---
that act on either end of a meson.
In the following  Section~\ref{s7}, we will talk about the algebra of the operators that
act on the gluons, hence in the middle portion of a meson state.  This will be the
centrix algebra. Finally, in  Section~\ref{s6}, we will piece together all the algebras we will have
studied so far, together with the finite-dimensional matrix algebras for
quarks and antiquarks, to form a grand Lie algebra.  

We discuss a couple of illlustrative examples of our algebra in the
last sections.
In Section~\ref{s9}, we will apply our algebra to study the relationship between 
quantum spin chain systems and matrix models.
In Section~\ref{s10}, we will show that QCD can be formulated in terms
of the grand algebra; we will also show to determine an approximate
spectrum analytically.

In Appendix~\ref{s1-1}, we will summarize the notations
used throughout this article.

\section{Physical States and Operators}
\label{s2}

We begin by introducing the basic microscopic degrees of freedom
of our theory, which satisfy  the Canonical (Anti-)Commutation Relations:
\begin{eqnarray}
   \left[ a^{\mu_1}_{\mu_2} (k_1), a^{\dagger\mu_3}_{\mu_4} (k_2) \right] & = & 
   \delta_{k_1 k_2} \delta^{\mu_1}_{\mu_4} \delta^{\mu_3}_{\mu_2}; \nonumber \\
   \left[ a^{\mu_1}_{\mu_2} (k_1), a^{\mu_3}_{\mu_4} (k_2) \right] & = & 0;
\nonumber \\
   \left[ a^{\dagger\mu_1}_{\mu_2} (k_1), a^{\dagger\mu_3}_{\mu_4} (k_2) \right] & = & 0; 
\label{1.0.0.1} \\
   \left[ q^{\mu_1} (k_1), q^{\dagger}_{\mu_2} (k_2) \right]_+ & = &
   \delta_{k_1 k_2} \delta^{\mu_1}_{\mu_2}; \nonumber \\
   \left[ q^{\mu_1} (k_1), q^{\mu_2} (k_2) \right]_+ & = & 0; \nonumber \\
   \left[ q^{\dagger}_{\mu_1} (k_1), q^{\dagger}_{\mu_2} (k_2) \right]_+ & = & 0; 
\label{1.0.0.2} \\
   \left[ \bar{q}_{\mu_1} (k_1), \bar{q}^{\dagger\mu_2} (k_2) \right]_+ & = &
   \delta_{k_1 k_2} \delta^{\mu_2}_{\mu_1}; \nonumber \\
   \left[ \bar{q}_{\mu_1} (k_1), \bar{q}_{\mu_2} (k_2) \right]_+ & = & 0
\mbox{; and} 
\nonumber \\
   \left[ \bar{q}^{\dagger\mu_1} (k_1), \bar{q}^{\dagger\mu_2} (k_2) \right]_+ & = & 0.  
\label{1.0.0.3}
\end{eqnarray} 
Here, \m{[A,B]=AB-BA} is a commutator and \m{[A,B]_+\equiv AB+BA} an anti-commutator.

The bosonic operators \m{a,a^{\dag}} create and annihilate fundamental
degrees of freedom that we will call `gluons'. This comes from the
application of our theory to  regularized  QCD, in which case these are gluon
operators. (But other interpretations in terms of string bits
\cite{thorn96}, D-branes \cite{dbrane} or M-theory \cite{susskind} are  also possible.) The indices
\m{\mu,\nu=1,2,\cdots, N} label a degree of freedom we will call
`color' following this analogy.  In the same way we will
call the states created by \m{q^{\dag}} `quarks' and \m{\bar{q}^{\dag}} `antiquarks'.
To simplify the following discussion, we will mostly  assume that gluons, quarks and 
antiquarks have only a finite number of distinct quantum states other than color. This can 
be done, for instance, by discretizing the momentum and other quantum numbers with 
continuous ranges of values.  Let $\Lambda$ be the possible number of
distinct quantum states of a gluon (not counting color), and let us use the numbers $1$, $2$, \ldots,
$\Lambda$ to denote these quantum states.  Likewise, let us use
$\lambda,\rho=1, 2, \ldots, \Lambda_F$ to label 
collectively the quantum numbers (other than color) of quarks or antiquarks.  

\begin{figure}
\epsfxsize=5in
\epsfysize=2.1in
\centerline{\epsfbox{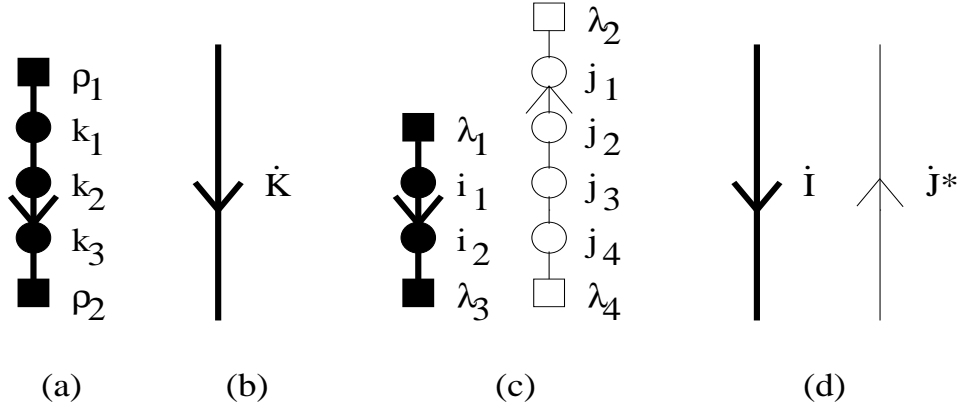}}
\caption{\em (a) A typical single meson state with 3 gluons in detail.  The solid square at 
the top is the creation operator of an antiquark of Quantum State $\rho_1$.  Following the 
antiquark is a series of 3 gluons of Quantum States $k_1$, $k_2$, and $k_3$ respectively.  
The creation operators of these gluons are represented by solid circles.  The solid square 
at the bottom is the creation operators of a quark of Quantum State $\rho_2$.  Note that all 
the creation operators carry color indices.  A solid line, no matter how thick it is, 
connecting two circles, or a circle and a square, is used to mean that the two corresponding 
creation operators share a color index, and this color index is being summed over.  The 
arrow indicates the direction of the integer sequence $\dot{K}$.  (b) A single meson state in 
brief.  The quark and aniquark are neglected.  They will be consistently ignored in all
brief diagrammatic representations.  The gluon series is represented by the 
integer sequence $\dot{K}$.  (c) A typical operator of the first kind.  The solid squares and 
circles are creation operators of a quark, an antiquark and gluons.  The hollow 
squares are annihilation operators of an antiquark of Quantum State $\lambda_2$ and a quark
of Quantum State $\lambda_4$, and the hollow circles are annihilation operators of gluons.  In this 
particular example, there are 2 creation and 4 annihilation operators of
gluons.  The creation operators are joined by thick lines, whereas the annihilation
operators are joined by thin lines.  Note that the sequence $\dot{J}$ is in reverse.  (d)  An operator of
the first kind in brief.  The thick line represents a sequence of creation operators of 
gluons, whereas the thin line represents a sequence of annihilation operators of them.  $\dot{J}$ carries an
asterisk to signify the fact that $\dot{J}$ is put in reverse.  Note 
that the lengths of the two lines have no bearing on the numbers of creation or annihilation 
operators they represent.}     
\label{f1}
\end{figure}

We are now ready to construct open string states, or meson states.  This is analogous to the way
closed string states, or glueball states, are constructed by Thorn \cite{th79}.
A meson consists of one quark, one antiquark and an arbitrary number of gluons.  Hence, a typical
colorless state of a meson can be written down as a linear combination of
\begin{eqnarray}
   \bar{\phi}^{\rho_1} \otimes s^{\dot{K}} \otimes \phi^{\rho_2} & \equiv & 
   N^{-(\dot{c}+1)/2} \bar{q}^{\dagger\upsilon_1} (\rho_1) a^{\dagger\upsilon_2}_{\upsilon_1}(k_1)
   \cdots \nonumber \\             
   & & a^{\dagger\upsilon_{\dot{c}+1}}_{\upsilon_{\dot{c}}} (k_{\dot{c}}) 
   q^{\dagger}_{\upsilon_{\dot{c}+1}} (\rho_2)    
   |0\rangle ,
\label{1.0.1}
\end{eqnarray}
where $\dot{c} = \#(\dot{K})$. Note that it is possible for \m{\dot{K}} to be 
empty, which corresponds to a meson state containing no gluons.  Moreover, 
the summation convention for color indices is implicit in the above and 
following expressions. The factor of $N^{-(\dot{c}+1)/2}$ is inserted so that the 
state has a finite norm in the large-$N$ limit.  The justification of the use of the symbol 
$\otimes$ on the left hand side will be given shortly.  A meson state given by Eq.(\ref{1.0.1}) is
depicted in Figs.~\ref{f1}(a) and (b). 

Let us turn our attention to operators representing dynamical
variables. As in models studied before (e.g., \cite{DaKl93a, AnDa,
thorn96}), the only operators we will look at are those which can propagate single meson
states to single meson states, in the leading order of the planar large-$N$
limit. For a finite value of \m{N},  they can also convert a 
single meson  to a state with more than one meson or glueball.  However, 
these terms are of order $1/N$ and so are suppresed in the planar large-$N$ limit.
This fact has been well known for a long time \cite{th79}, and is closely 
related to the planarity of Feynman diagrams of perturbation theory \cite{plan}.  A non-rigorous
diagrammatic proof of this fact is given in Ref.\cite{lee}. 

There are four kinds of such gauge invariant operators that survive in the large-$N$ limit.  
The simplest, {\em operators of the first kind} are {\em finite} linear combinations of operators of the form
\begin{eqnarray}
   \bar{\Xi}^{\lambda_1}_{\lambda_2} \otimes f^{\dot{I}}_{\dot{J}} \otimes \Xi^{\lambda_3}_{\lambda_4} 
   & \equiv & N^{-(\dot{a}+\dot{b}+2)/2} \bar{q}^{\dagger\mu_1} (\lambda_1) 
   a^{\dagger\mu_2}_{\mu_1}(i_1) \cdots a^{\dagger\mu_{\dot{a}+1}}_{\mu_{\dot{a}}} (i_{\dot{a}}) 
   q^{\dagger}_{\mu_{\dot{a}+1}} (\lambda_3) \cdot \nonumber \\
   & & \bar{q}_{\nu_1} (\lambda_2) a^{\nu_1}_{\nu_2} (j_1) \cdots 
   a^{\nu_{\dot{b}}}_{\nu_{\dot{b}+1}} (j_{\dot{b}}) q^{\nu_{\dot{b}+1}} (\lambda_4), 
\label{1.0.2}
\end{eqnarray}
where $\dot{a} = \#(\dot{I})$ and $\dot{b} = \#(\dot{J})$.  In the planar limit, an operator of the
first kind propagates a single meson state to either zero or another single meson state:
\begin{equation}
   \bar{\Xi}^{\lambda_1}_{\lambda_2} \otimes f^{\dot{I}}_{\dot{J}} \otimes \Xi^{\lambda_3}_{\lambda_4}
   \left( \bar{\phi}^{\rho_1} \otimes s^{\dot{K}} \otimes \phi^{\rho_2} \right) =
   \delta^{\rho_1}_{\lambda_2} \delta^{\dot{K}}_{\dot{J}} \delta_{\lambda_4}^{\rho_2}
   \bar{\phi}^{\lambda_1} \otimes s^{\dot{I}} \otimes \phi^{\lambda_3}.
\label{2.0.0.1}
\end{equation}

It is now clear why we are using the direct product symbol $\otimes$ ---  the single meson state 
can be regarded as a direct product of $\bar{\phi}^{\rho_1}$, $s^{\dot{K}}$ 
and $\phi^{\rho_2}$.  The set of all $\bar{\phi}^{\rho_1}$'s, where $\rho_1 = 1, 2$, \ldots, and 
$\Lambda_F$, form a basis of a $\Lambda_F$-dimensional vector space.  The set of all $\phi^{\rho_2}$'s, 
where again $\rho_2 = 1, 2$, \ldots, and $\Lambda_F$, form a basis of another $\Lambda_F$-dimensional 
vector space.  The operator of the first kind can be regarded as a direct product of the
operators $\bar{\Xi}^{\lambda_1}_{\lambda_2}$, $f^I_J$ and $\Xi^{\lambda_3}_{\lambda_4}$.  The first
operator acts as a $\Lambda_F \times \Lambda_F$ matrix on $\bar{\phi}^{\rho_1}$, the
second one acts on $s^{\dot{K}}$, whereas the last one acts as another $\Lambda_F \times \Lambda_F$
matrix on $\phi^{\rho_2}$.  It is therefore clear that that a meson state lies
within a direct product of two $\Lambda_F$-dimensional vector spaces (labelling the quark states)  and a
countably infinite-dimensional vector space spanned by all $s^{\dot{K}}$'s labelling the
gluon states ( including the state containing no gluons).  An operator of the first kind lies within the
direct product $gl(\Lambda_F) \otimes \salt \otimes gl(\Lambda_F)$. Here, $gl(\Lambda_F)$
is the Lie algebra of the general linear group $GL(\Lambda_F)$. 
Also the infinite-dimensional Lie algebra  $\salt$ is spanned by 
\m{f^{\dot{I}}_{\dot{J}}}. We will see that \m{\salt}  is isomorphic to the
inductive limit \m{gl_{+\infty}} of the \m{gl(n)}'s as \m{n\to \infty}.

{\em Operators of the second kind} are {\em finite} linear combinations of 
operators of the form
\begin{eqnarray}
   \bar{\Xi}^{\lambda_1}_{\lambda_2} \otimes l^{\dot{I}}_{\dot{J}} \otimes 1 & \equiv & 
   N^{-(\dot{a}+\dot{b})/2} \bar{q}^{\dagger\mu_1} (\lambda_1) 
   a^{\dagger\mu_2}_{\mu_1} (i_1) a^{\dagger\mu_3}_{\mu_2} (i_2) \cdots
   a^{\dagger\mu_{\dot{a}+1}}_{\mu_{\dot{a}}} (i_{\dot{a}}) \nonumber \\
   & & a^{\nu_{\dot{b}}}_{\mu_{\dot{a}+1}} (j_{\dot{b}}) 
   a^{\nu_{\dot{b}-1}}_{\nu_{\dot{b}}} (j_{\dot{b}-1}) \cdots
   a^{\nu_1}_{\nu_2} (j_1) \bar{q}_{\nu_1} (\lambda_2)
\label{1.1}
\end{eqnarray}
whereas an {\em operator of the third kind} can be written as a linear combination of operators of the form   
\begin{eqnarray}
   1 \otimes r^{\dot{I}}_{\dot{J}} \otimes \Xi^{\lambda_1}_{\lambda_2} & \equiv & 
   N^{-(\dot{a}+\dot{b})/2} q^{\dagger}_{\mu_{\dot{a}+1}} (\lambda_1) 
   a^{\dagger\mu_{\dot{a}+1}}_{\mu_{\dot{a}}} (i_{\dot{a}}) 
   a^{\dagger\mu_{\dot{a}}}_{\mu_{\dot{a}-1}} (i_{\dot{a}-1}) \cdots \nonumber \\
   & & a^{\dagger\mu_2}_{\mu_1} (i_1) a^{\mu_1}_{\nu_1} (j_1) a^{\nu_1}_{\nu_2} 
   (j_2) \cdots a^{\nu_{\dot{b}-1}}_{\nu_{\dot{b}}} (j_{\dot{b}}) q^{\nu_{\dot{b}}} (\lambda_2).
\label{1.2}
\end{eqnarray}

Both the second and third kinds are regularized versions of the
`current density' operators of QCD.  There are two components to the
currents, namely the left-handed component ${\bar q}^a(x){\bar q}^b(x)$ and right-handed component 
$q^a (x) q^b (x)$, where \m{a,b} are spin and
flavor indices. (They act on the left and right ends respectively of a meson.)
 These operators however  are not well defined due to the
divergences of quantum field theory. We must instead consider the
point-split form, where the fields are evaluated at different points
in space. But then gauge invariance forces us to insert a parallel
tranport operator along a path \m{P} that connects the two points. Then
the current operators become functions of an open path, i.e.,  string-like
operators:
\beq
	\bar q^a(x)e^{i\int_P Adx}\bar q^b(y), \quad
				 q^a(x)e^{i\int_P Adx}q^b(y).
\eeq
We can  now expand the field in terms of creation--annihilation
operators. Also, we must introduce a regularization that allows only a
finite number of values for the momenta of quarks and gluons, in order
to get a rigorous theory. These are precisly our operators  of the
first three kinds. What we discover is that in the large \m{N}
limit these operators form a closed Lie algebra. Thus we have found the
current algebra for QCD in the planar large-\m{N} limit.

Let us see how these operators act on a meson state, in the  large
 \m{N} limit.
 Foe the second kind,
\begin{equation}
   \bar{\Xi}^{\lambda_1}_{\lambda_2} \otimes l^{\dot{I}}_{\dot{J}} \otimes 1 \left( 
   \bar{\phi}^{\rho_1} \otimes s^{\dot{K}} \otimes \phi^{\rho_2} \right) =
   \delta^{\rho_1}_{\lambda_2} \sum_{\dot{K}_1\dot{K}_2 = \dot{K}} 	
   \delta^{\dot{K}_1}_{\dot{J}} \bar{\phi}^{\lambda_1} \otimes s^{\dot{I} \dot{K}_2} \otimes \phi^{\rho_2}. 
\label{2.0.1}
\end{equation}
There will only be a finite number of non-zero terms in this sum (bounded by the
number of ways of splitting \m{\dot{K}} into sub-sequences), so
there are no problems of convergence.
For example, if \m{\dot{K}} is shorter than \m{\dot{J}}, the
right hand side will be zero. 
The action of an operator of the third kind is similar except that it acts on
the quark end:
\begin{equation}
   1 \otimes r^{\dot{I}}_{\dot{J}} \otimes \Xi^{\lambda_1}_{\lambda_2} \left(
   \bar{\phi}^{\rho_1} \otimes s^{\dot{K}} \otimes \phi^{\rho_2} \right) = 
   \delta_{\lambda_2}^{\rho_2} \sum_{\dot{K}_1 \dot{K}_2 = \dot{K}} 		
   \delta^{\dot{K}_2}_{\dot{J}} \bar{\phi}^{\rho_1} \otimes s^{\dot{K}_1
   \dot{I}} \otimes \phi^{\lambda_1}.
\label{2.0.2}
\end{equation}

We saw that a properly regularized version of the current operators
invloves the gluon field: This means that we must also consider
analoges of current densities that are made up entirely of gluons. In
any case such terms will be present in the hamiltonian of QCD.
They  will dominate the dynamics of QCD in the  large-\m{N}
limit. These lead us to 
{\em operators of the fourth kind}, or \emph{gluonic operators}, which are finite linear combinations of operators
of the form:
\begin{eqnarray}
   1 \otimes \s^I_J \otimes 1 & \equiv & N^{-(a+b-2)/2} a^{\dagger\mu_2}_{\mu_1} (i_1)
   a^{\dagger\mu_3}_{\mu_2} (i_2) \cdots a^{\dagger\nu_b}_{\mu_a} (i_a) 
   \nonumber \\
   & & a^{\nu_{b-1}}_{\nu_b} (j_b) a^{\nu_{b-2}}_{\nu_{b-1}} (j_{b-1}) \cdots
   a^{\mu_1}_{\nu_1} (j_1),
\label{1.3}
\end{eqnarray}
where $a = \#(I)$ and $b = \#(J)$.
Unlike the operators of the first three kinds, \emph{both $I$ and $J$ in the
operators of the fourth kind must be non-empty sequences}.  
In the large-$N$ limit, a gluonic operator propagates a single meson state to a
linear combination of single meson states:
\begin{equation}
   1 \otimes \s^I_J \otimes 1 \left( \bar{\phi}^{\rho_1} \otimes s^{\dot{K}} \otimes \phi^{\rho_2} \right) = 
   \bar{\phi}^{\rho_1} \otimes \left( \sum_{\dot{K}_1 K_2 \dot{K}_3 = \dot{K}} \delta^{K_2}_{J}
   s^{\dot{K_1} I \dot{K_3}} \right) \otimes \phi^{\rho_2}.
\label{1.4}
\end{equation} 
We see clearly from this equation that the gluonic operator acts trivially on
the quark and antiquark.  It acts on a central portion of a meson, i.e., the gluons 
lying in between the quark/antiquark pair, or on a gluon segment of a
glueball.

\begin{figure}
\epsfxsize=5.5in
\epsfysize=4in
\centerline{\epsfbox{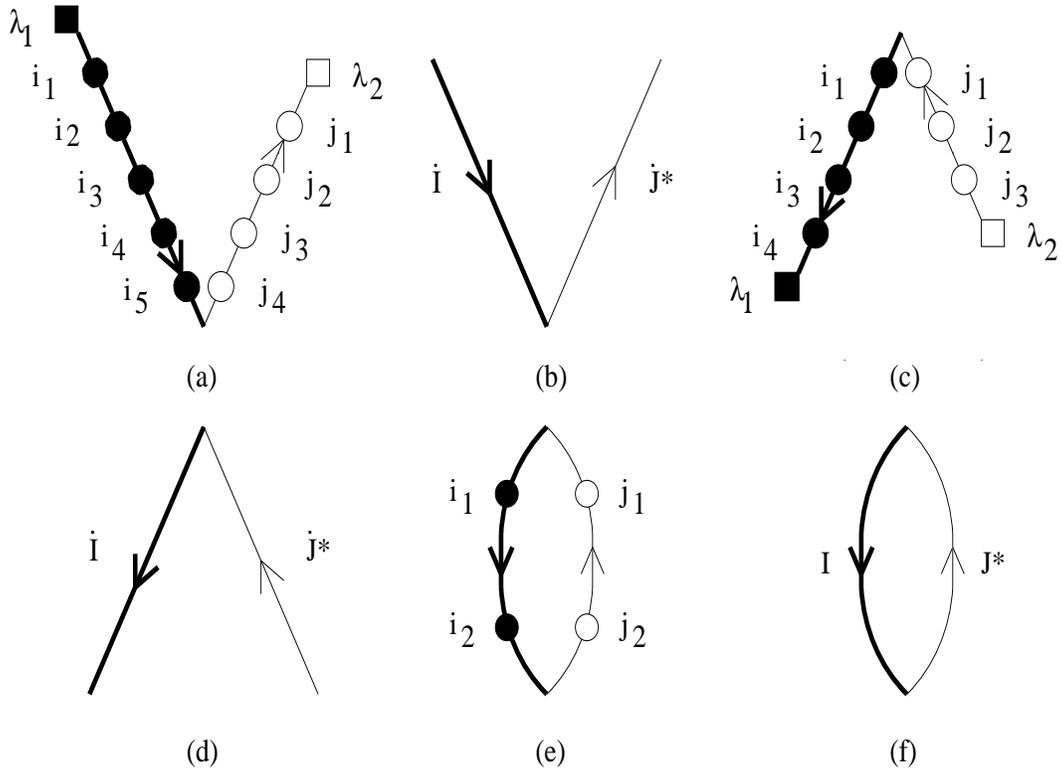}}
\caption{\em  (a) A typical operator of the second kind acting on the antiquark end in detail.  There
are 5 creation and 4 annihilation operators of gluons.  The solid square is a creation 
operator of an antiquark, whereas the hollow square is an annihilation operator of it.
(b) An operator of the second kind acting on the antiquark end in brief.
(c) A typical operator of the third kind acting on the quark end in detail.  There are 4
creation and 3 annihilation operators of gluons in this current operator.  This time the 
solid square is a creation operator of a quark, whereas the hollow square is an
annihilation operator of it.  Compare the orientations of the arrows with those in (a).  The choices of the
orientations reflect the fact that the color indices are contracted differently in Eqs.(\ref{1.1}) and (\ref{1.2}).
(d) An operator of the third kind acting on the quark end in brief.  (e) A
typical gluonic operator in detail.  There are 2 creation and 2 annihilation operators
of gluons.  There are no operators acting on a quark or an antiquark.  (f) A gluonic operator
in brief.}
\label{f2}
\end{figure}

The operators defined in Eqs.(\ref{1.1}), (\ref{1.2}) and (\ref{1.3}) are
depicted in Figs.~\ref{f1}(c), (d) and \ref{f2}.

As can be seen from the four kinds of operators, the essentially new algebraic
structures are contained in the gluonic indices.  In the succeeding sections, we are going to confine
ourselves to the gluonic parts of the physical states and operators.  The quark degrees of freedom will
be put back later when we discuss the overall algebra, or the `grand' algebra. 

\section{Current Algebra in the Large-$N$ Limit}
\label{s3}

In this section, we are going to study the actions of the operators of the first
three kinds on the gluon segment of a meson. This can be viewed as providing a
representation of an associative  algebra among these operators.  We will see
that the algebra of the first kind is a direct product of 2
finite-dimensional algebras and a well-known  infinite-dimensional 
algebra, $gl_{+\infty}$ \cite{8, KaPe}: it is thus not an essentially new object.  

Then we will turn our attention to current operators; i.e, those of the second and third kinds.  
We will see that the associative algebra among the current  operators is a direct
product of a  finite-dimensional algebra and an infinite-dimensional associative algebra closely
related to the Cuntz algebra \cite{3, 4}.  Then we will focus on this 
infinite-dimensional algebra and study its structure. We will 
particularly be interested in the Lie algebra defined by the commutator
in this new infinite-dimensional Lie algebra: it is a kind  of 
current algebra in our theory. 
We will relate this current algebra with 
the Cuntz algebra \cite{7} and Kac-Moody algebra \cite{8}. 

\begin{figure}[ht]
\epsfxsize=5in
\centerline{\epsfbox{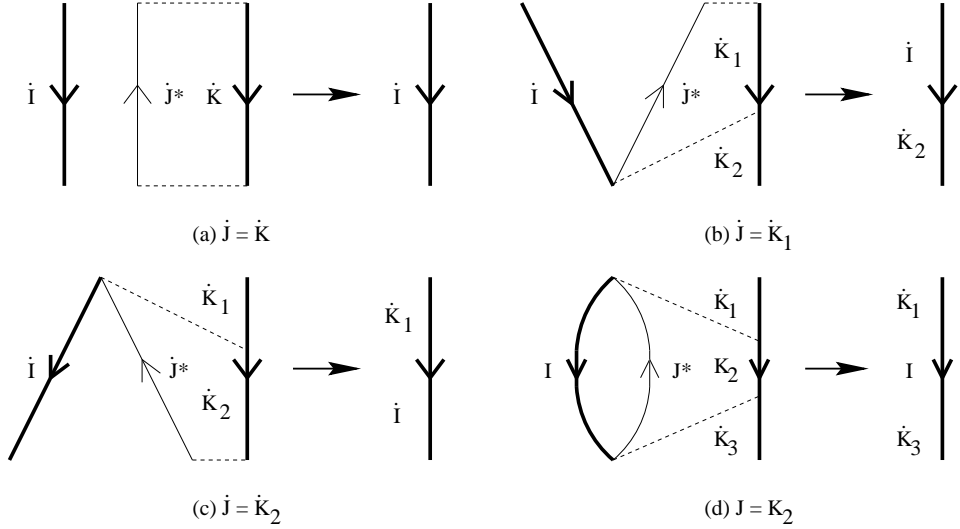}}
\caption{\em (a) The action of an operator of the first kind on the gluon portion of a single meson state 
(Eq.(\ref{3.3.1})).  The dotted lines connect the line segments to be `annihilated' together.  The figure 
on the right of the arrow is the resultant meson state.  (b) and (c) The actions of current operators on
the gluon portion of a single meson state.  In (b) we have the diagrammatic representation of the R.H.S. 
of Eq.(\ref{2.1}), whereas in (c) we have the representation of the R.H.S. of Eq.(\ref{2.1.1}).  
(d)  The action of a gluonic operator on a single meson state.} 
\label{f3}
\end{figure}

Let us consider operators of the first kind but ignoring the action on
the quark indices; this corresponds to setting \m{\Lambda_F=1} temporarily. Then Eq.(\ref{2.0.0.1})
tells us that
\begin{equation}
   f^{\dot{I}}_{\dot{J}} s^{\dot{K}} = \delta^{\dot{K}}_{\dot{J}} s^{\dot{I}}.
\label{3.3.1}
\end{equation}
This equation is diagrammatically illustrated in Fig.~\ref{f3}(a).  In
particular, $f^{\emptyset}_{\emptyset}$, where \m{\emptyset} is the null 
sequence, is nothing but the projection operator to the gluonless state.  Define the composite operator
$f^{\dot{I}}_{\dot{J}} f^{\dot{K}}_{\dot{L}}$ by the requirement that 
its action on $s^M$ give $f^{\dot{I}}_{\dot{J}} (f^{\dot{K}}_{\dot{L}} s^{\dot{M}})$.   Then it follows that
\begin{equation}
   f^{\dot{I}}_{\dot{J}} f^{\dot{K}}_{\dot{L}} = \delta^{\dot{K}}_{\dot{J}} f^{\dot{I}}_{\dot{L}}.
\end{equation}
Thus the $f$'s form an algebra.  Moreover, it is obvious that this
algebra is the associative algebra of finite-rank matrices on the
infinite dimensional space spanned by the \m{s^{\dot{I}}}.
The \m{ f^{\dot{I}}_{\dot{J}}} are Weyl matrices in this orthonormal basis.

Let us define a Lie algebra \m{\salt} spanned by the \m{ f^{\dot{I}}_{\dot{J}} }
,  with the usual commutator.  Then 
\begin{equation}
   \left[ f^{\dot{I}}_{\dot{J}}, f^{\dot{K}}_{\dot{L}} \right] = 
   \delta^{\dot{K}}_{\dot{J}} f^{\dot{I}}_{\dot{L}} - \delta^{\dot{I}}_{\dot{L}} f^{\dot{K}}_{\dot{J}}.
\label{3.6.1}
\end{equation}
Hence it is clear that ${\salt}$ is isomorphic to $gl_{+\infty}$, which is
the Lie algebra, with the usual bracket, of all complex matrices 
$(a_{ij})_{i, j} \in Z_+$ such that the number of nonzero $a_{ij}$'s is finite \cite{8}.
The isomorphism is given by a one-one correspondence between the multi-indices
\m{I} and the set of natural numbers \m{Z_+}. 

Let us now turn to current operators.  Ignoring the action on the (anti)-quark indices,  
we can simplify Eqs.(\ref{2.0.1}) and (\ref{2.0.2}) to
\begin{eqnarray}
   l^{\dot{I}}_{\dot{J}} s^{\dot{K}} & = & 
   \sum_{\dot{K}_1 \dot{K}_2 = \dot{K}} \delta^{\dot{K}_1}_{\dot{J}} s^{\dot{I} \dot{K}_2} \mbox{; and} 
\label{2.1} \\
   r^{\dot{I}}_{\dot{J}} s^{\dot{K}} & = & 
   \sum_{\dot{K}_1 \dot{K}_2 = \dot{K}} \delta^{K_2}_{\dot{J}} s^{\dot{K}_1 \dot{I}},   
\label{2.1.1}
\end{eqnarray}
The reader can appreciate the meanings of these equations better by looking at Fig.~\ref{f3}(b) and (c).  
Note that $l^{\emptyset}_{\emptyset}=r^{\emptyset}_{\emptyset}=1$, the identity operator.  

Let us consider the multiplication rule for the $l^{\dot{I}}_{\dot{J}}$'s.  The properties of the algebra 
of the set of all $r^{\dot{I}}_{\dot{J}}$'s are completely analogous.
Let us define the composite operator $l^{\dot{I}}_{\dot{J}} l^{\dot{K}}_{\dot{L}}$ by the 
requirement that its action on $s^{\dot{M}}$ give $l^{\dot{I}}_{\dot{J}} (l^{\dot{K}}_{\dot{L}} s^{\dot{M}})$.  
Then it follows that
\begin{equation}
   l^{\dot{I}}_{\dot{J}} l^{\dot{K}}_{\dot{L}} = \delta^{\dot{K}}_{\dot{J}} l^{\dot{I}}_{\dot{L}} +
   \sum_{\dot{J}_1 J_2 = \dot{J}} \delta^{\dot{K}}_{\dot{J}_1} l^{\dot{I}}_{\dot{L} J_2} + 
   \sum_{\dot{K}_1 K_2 = \dot{K}} \delta^{\dot{K}_1}_{\dot{J}} l^{\dot{I} K_2}_{\dot{L}}.
\label{2.1.2}
\end{equation}
Here we sum over all possible sequences except that \m{J_2} and \m{K_2} are
required to be non-empty.  \m{\dot{J_2}} being empty and \m{\dot{K_2}} being empty describe 
really the first term on the R.H.S. and we want to make sure that this term is counted only once in the sums. 

Thus the $l$'s form an algebra.  This algebra is associative \cite{lee}.  Moreover, this algebra has an intimate
relationship with the extended Cuntz algebra \cite{3, 4}.  Indeed, let ${\cal H}$ be a $\Lambda$-dimensional 
Hilbert space spanned by $v^1$, $v^2$, \ldots, $v^{\Lambda}$, and ${\cal F} ({\cal H})$ be the full Fock space 
$\oplus_{r=0}^{\infty} (\otimes^r {\cal H})$, where $(\otimes^0 {\cal H})$ is a
one-dimensional Hilbert space spanned by a unit vector $\Omega$, the `vacuum'
\footnote{For us the state \m{\Omega} is not exactly the vacuum, but rather is  the gluonless state,
which still contains a quark and an antiquark.}.  We will see that ${\cal F} ({\cal H})$ is a representation space 
for the extended Cuntz algebra.  There is a bijective mapping between the Fock space of physical states spanned by 
single mesons and a vector in the space ${\cal F} ({\cal H})$.  The state $s^{\emptyset}$ is mapped 
to the $\Omega$, and the state $s^K$ is mapped to 
$v^{k_1} \otimes v^{k_2} \otimes \cdots \otimes v^{k_c}$.  Define the operators
$a^{1\dagger}$, $a^{2\dagger}$, \ldots, $a^{{\Lambda} \dagger}$ as follows:
\begin{eqnarray}
   a^{i\dagger} v^{k_1} \otimes v^{k_2} \otimes \cdots \otimes v^{k_c} & = &
   v^i \otimes v^{k_1} \otimes v^{k_2} \otimes \cdots \otimes v^{k_c} 
   \nonumber \\
   a^{i\dagger} \Omega & = & v^i
\end{eqnarray}
where $i$ is an integer between $1$ and $\Lambda$ inclusive.  The corresponding 
adjoint operators $a_1$, $a_2$, \ldots, $a_{{\Lambda}}$ have the following properties:
\begin{eqnarray}
   a_i v^{k_1} \otimes v^{k_2} \otimes \cdots \otimes v^{k_c} & = &
   \delta^{k_1}_i v^{k_2} \otimes \cdots \otimes v^{k_c} \mbox{; and} 
   \nonumber \\
   a_i \Omega & = & 0.
\end{eqnarray}
As a result, the $a$'s and $a^{\dagger}$'s satisfy the following properties:
\begin{eqnarray}
   a_i a^{j\dagger} & = & \delta^j_i \mbox{; and} \\
   \sum_{i=1}^{{\Lambda}} a^{i\dagger} a_i & = & 1 - P_\Omega ,
\label{4.1}
\end{eqnarray}
where $P_{\Omega}$ is the projection operator to the vacuum $\Omega$.  Thus the
$a$'s and $a^{\dagger}$'s are the annihilation and creation operators of the 
extended Cuntz algebra respectively.  Furthermore,
\begin{equation}
   v^{k_1} \otimes v^{k_2} \otimes \cdots \otimes v^{k_c} =
   a^{k_1 \dagger} a^{k_2 \dagger} \cdots a^{k_c \dagger} \Omega .
\end{equation}

It is now straightforward to see that there is a one-to-one correspondence among 
the operators characterized by Eq.(\ref{1.1}) and the operators acting on ${\cal F} 
({\cal H})$.  A current operator $l^I_J$ corresponds to 
$a^{i_1 \dagger} a^{i_2 \dagger} \cdots a^{i_a \dagger} a_{j_b} a_{j_{b-1}} 
\cdots a_1$;$l^I_{\emptyset}$ corresponds to $a^{i_1 \dagger} a^{i_2 \dagger} \cdots
a^{i_a \dagger}$; $l^{\emptyset}_J$ corresponds to $a_{j_b} a_{j_{b-1}} 
\cdots a_1$; and $l^\emptyset_\emptyset$ to the identity operator.

\begin{figure}[ht]
\epsfxsize=5in
\centerline{\epsfbox{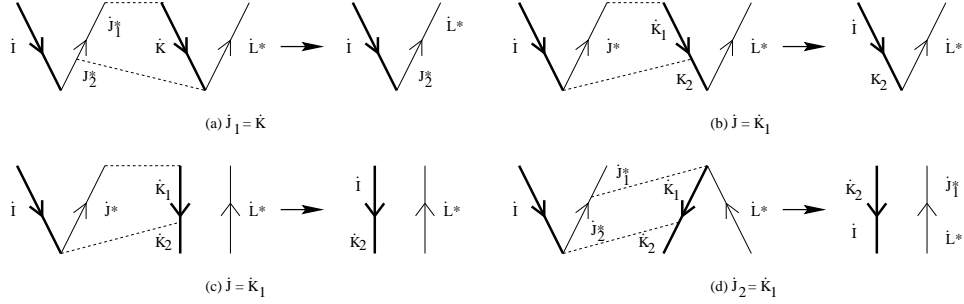}}
\caption{\em (a) and (b) Diagrammatic representations of the defining Lie bracket of the leftix algebra.  (a) shows
the first two terms whereas (b) shows the third term of the R.H.S. of Eq.(\ref{2.2}) are shown.  (c) 
The Lie bracket between $l^{\dot{I}}_{\dot{J}}$ and $f^{\dot{K}}_{\dot{L}}$.  Only the first term on the R.H.S.
of Eq.(\ref{3.5}) is shown.  (d) The Lie bracket between $l^{\dot{I}}_{\dot{J}}$ and $r^{\dot{K}}_{\dot{L}}$.  Only
the first term on the R.H.S. of Eq.(\ref{7.8.2}) is shown.}
\label{f4}
\end{figure}

Define the Lie algebra between the $l$'s as follows:
\begin{eqnarray}
   \lbrack l^{\dot{I}}_{\dot{J}}, l^{\dot{K}}_{\dot{L}} \rbrack & = & 
   \delta^{\dot{K}}_{\dot{J}} l^{\dot{I}}_{\dot{L}} +
   \sum_{\dot{J}_1 J_2 = \dot{J}} \delta^{\dot{K}}_{\dot{J}_1} l^{\dot{I}}_{\dot{L} J_2} + 
   \sum_{\dot{K}_1 K_2 = \dot{K}} \delta^{\dot{K}_1}_{\dot{J}} l^{\dot{I} K_2}_{\dot{L}} \nonumber \\
   & & - \delta^{\dot{I}}_{\dot{L}} l^{\dot{K}}_{\dot{J}} -
   \sum_{\dot{L}_1 L_2 = \dot{L}} \delta^{\dot{I}}_{\dot{L}_1} l^{\dot{K}}_{\dot{J} L_2} - 
   \sum_{\dot{I}_1 I_2 = \dot{I}} \delta^{\dot{I}_1}_{\dot{L}} l^{\dot{K} I_2}_{\dot{J}}.
\label{2.2}
\end{eqnarray}
The first three terms on the R.H.S. of Eq.(\ref{2.2}) are diagrammatically 
represented in Fig.~\ref{f4}(a) and (b).  Let us call the Lie algebra defined by Eq.(\ref{2.2}) the 
{\em leftix algebra} or $\hatleftix$. (We justify this name as an abbreviation of
`left multi-matrix'.) The analogous algebra spanned by \m{r^{\dot{I}}_{\dot{J}}} will
be called the rightix algebra $\hatrightix$.

Now we note the following identity:
\begin{equation}
    \left( l^{\dot{K}}_{\dot{L}} - \sum_{i=1}^{\Lambda} l^{\dot{K}i}_{\dot{L}i} \right) s^{\dot{M}} = 
    \delta^{\dot{M}}_{\dot{L}} s^{\dot{K}}.
\end{equation}
Thus these are  nothing but $f^{\dot{K}}_{\dot{L}}$ we saw earlier:
\begin{equation}
   f^{\dot{K}}_{\dot{L}} \equiv l^{\dot{K}}_{\dot{L}} - \sum_{j=1}^{\Lambda} l^{\dot{K}j}_{\dot{L}j}.
\label{3.3}
\end{equation}
We can obtain from Eqs.(\ref{3.3.1}) and (\ref{2.1}) the following relation:
\begin{equation}
   \left[ l^{\dot{I}}_{\dot{J}}, f^{\dot{K}}_{\dot{L}} \right] = 
   \sum_{\dot{K}_1 \dot{K}_2 = \dot{K}} \delta^{\dot{K}_1}_{\dot{J}} f^{\dot{I}\dot{K}_2}_{\dot{L}}
   - \sum_{\dot{L}_1 \dot{L}_2 = \dot{L}} \delta^{\dot{I}}_{\dot{L}_1} f^{\dot{K}}_{\dot{J} \dot{L}_2}.
\label{3.5}
\end{equation}
This formula is depicted in Figs.~\ref{f4}(c).

As a result of Eq.(\ref{3.5}), ${\salt}$ forms an ideal of the algebra
$\hatleftix$.  This is a proper ideal as it is obvious that {\em
finite} linear combinations of \m{f^{\dot{I}}_{\dot{J}}}  do not span the whole
leftix algebra.  The quotient \m{\leftix=\hatleftix/\salt} is thus also a Lie algebra.
Put it another way, $\hatleftix$ is the extension of $\leftix$ by $\salt$:
we have the exact sequence of Lie algebras
\beq
	0\to \salt\to\hatleftix\to\leftix\to 0.
\eeq
At the moment we have only this indirect construction of the Lie
algebra $\leftix$.  It would be exciting to find new representations
for this leftix algebra.

How do we understand the above exact sequence in the context of the Cuntz algebra?  
The analogue of the algebra $F_{\Lambda}$ is the algebra ${\cal K} ({\cal H})$ of {\em compact} operators
on ${\cal H} = {\bf C}^{\Lambda}$; ${\cal K} ({\cal H})$ is just the completion of $F_{\Lambda}$ in the
topology defined by the operator norm.  It follows that ${\cal K} ({\cal H})$ is
an ideal of the Banach algebra generated by \m{a,a^{\dag}}.  
If we quotient  by ${\cal K} ({\cal H})$, then we get the Cuntz 
algebra, with the $a$'s and $a^{\dagger}$'s satisfying the following
relations:
\begin{eqnarray}
   a_i a^{j\dagger} & = & \delta^j_i \mbox{; and} \\
   \sum_{i=1}^{{\Lambda}} a^{i\dagger} a_i & = & 1.
\label{3.6}
\end{eqnarray}
The role of the last relation is to set all \m{f^{\dot{I}}_{\dot{J}}}'s
to zero; i.e., to quotient by $\salt$.  Thus, we can regard the Lie algebra $\hatleftix$ as an extension of the Lie
algebra associated with the Cuntz algebra by the subalgebra ${\salt}$.

We will see that in fact the presence of \m{\salt} as a proper ideal is a generic feature of most of the 
algebras we will study. Quotienting by this ideal will get us the  essentially new
algebras we are interested in studying. But it is only the extension
that will have interesting representations.  It appears that the extension by $gl_{+\infty}$ plays a role in our
theory that central extensions play in the theory of Kac-Moody and
Virasoro algebras. Similar extensions have appeared in previous
approaches to current algebras \cite{MiRa}.

There is of course a parallel theory for $\hatrightix$ which is a Lie algebra spanned by elements of the form
$r^{\dot{I}}_{\dot{J}}$, and $\rightix=\hatrightix/\salt$ which is the quotient of $\hatrightix$ by $\salt$.  
Let us briefly describe it here for the sake of completeness. The Lie bracket expression between two $r$'s is:
\begin{eqnarray}
   \lbrack r^{\dot{I}}_{\dot{J}}, r^{\dot{K}}_{\dot{L}} \rbrack & = & 
   \delta^{\dot{K}}_{\dot{J}} r^{\dot{I}}_{\dot{L}} +
   \sum_{J_1 \dot{J}_2 = \dot{J}} \delta^{\dot{K}}_{\dot{J}_2} l^{\dot{I}}_{J_1 \dot{L}} + 
   \sum_{K_1 \dot{K}_2 = \dot{K}} \delta^{\dot{K}_2}_{\dot{J}} l^{K_1 \dot{I}}_{\dot{L}} \nonumber \\
   & & - \delta^{\dot{I}}_{\dot{L}} l^{\dot{K}}_{\dot{J}} -
   \sum_{L_1 \dot{L}_2 = \dot{L}} \delta^{\dot{I}}_{\dot{L}_2} l^{\dot{K}}_{L_1 \dot{J}} - 
   \sum_{I_1 \dot{I}_2 = \dot{I}} \delta^{\dot{I}_2}_{\dot{L}} l^{I_1 \dot{K}}_{\dot{J}}.
\label{3.9}
\end{eqnarray}
A vector in $\salt$ can be expressed as an element in $\hatrightix$ as well:
\begin{equation}
   f^{\dot{K}}_{\dot{L}} = r^{\dot{K}}_{\dot{L}} - \sum_{i=1}^{\Lambda} r^{i\dot{K}}_{i\dot{L}}.
\label{33}
\end{equation}
The Lie bracket between an $r$ and an $f$ is
\begin{equation}
   \left[ r^{\dot{I}}_{\dot{J}}, f^{\dot{K}}_{\dot{L}} \right] = 
   \sum_{\dot{K}_1 \dot{K}_2 = \dot{K}} \delta^{\dot{K}_2}_{\dot{J}} f^{\dot{K}_1 \dot{I}}_{\dot{L}}
   - \sum_{\dot{L}_1 \dot{L}_2 = \dot{L}} \delta^{\dot{I}}_{\dot{L}_2} f^{\dot{K}}_{\dot{L}_1 \dot{J}}.
\label{3.8}
\end{equation}

The algebra $\hatmultix$ spanned by the operators \m{l^I_J,r^I_J} together is also interesting.  $\hatleftix$ and
$\hatrightix$ are subalgebras of $\hatmultix$.  In addition, the Lie bracket between an element of $\hatleftix$ and
an element of $\hatrightix$ is
\begin{equation}
   \left[ l^{\dot{I}}_{\dot{J}}, r^{\dot{K}}_{\dot{L}} \right] = 
   \sum_{\begin{array}{l}
   	    \dot{J}_1 \dot{J}_2 = \dot{J} \\
   	    \dot{K}_1 \dot{K}_2 = \dot{K}
   	 \end{array}} 
   \delta^{\dot{K}_1}_{\dot{J}_2} f^{\dot{I} \dot{K}_2}_{\dot{J}_1 \dot{L}} -
   \sum_{\begin{array}{l}
   	    \dot{I}_1 \dot{I}_2 = \dot{I} \\
   	    \dot{L}_1 \dot{L}_2 = \dot{L}
   	 \end{array}}
   \delta^{\dot{I}_2}_{\dot{L}_1} f^{\dot{I}_1 \dot{K}}_{\dot{J} \dot{L}_2}.
\label{7.8.2}
\end{equation}
A heuristic way to see that Eq.(\ref{7.8.2}) is ture is to verify that the action of the R.H.S. on an arbitrary
state $s^{\dot{M}}$ gives $l^{\dot{I}}_{\dot{J}} (r^{\dot{K}}_{\dot{L}} s^{\dot{M}}) - 
r^{\dot{K}}_{\dot{L}} (l^{\dot{I}}_{\dot{J}} s^{\dot{M}})$.  This, however, does not automatically lead to the
conclusion that the Jacobi identity is satisfied by this definition of the Lie bracket (Eq.(\ref{7.8.2})) because
Eq.(\ref{3.3}) and (\ref{33}) imply that the set of all finite linear combinations of $l$'s and $r$'s is not
linearly independent.  The reader is referred to a future publication \cite{sustal} for a rigorous proof of 
Eq.(\ref{7.8.2}).  A typical term in the above equation is depicted in Fig.~\ref{f4}(d).

Therefore $\hatleftix$ and $\hatrightix$ are proper ideals of $\hatmultix$.  In particular, $\salt$ is a proper
ideal of $\hatmultix$, hence we have the quotient algebra
$\multix=\hatmultix/\salt$. Since the commutator of an \m{l} and an \m{r} is in \m{\salt}, they
commute when thought of as elements of \m{\multix}.  

These operators are regularized versions of the left- and right-handed current operators of
QCD. We see that the chiral structure of the current algebra  is
reproduced beautifully in the multix algebra. The fact that only the
extension of the current algebra  by \m{\gl} has a representation on
the one-meson states is also reminiscent of what happens in lower
dimensional theories.

Along the way we have learned an important lesson: the precisley defined
left and right currents {\em do not  commute}: their commutator is a
finite rank operator instead. Our naive expectations on current
algebra have to be
revised: our extensions by \m{\salt} are just as important as the
central extensions in the theory of Kac--Moody algebras.

We will see that much of this structure is embedded in the algebra of gluonic operators on open string states. We 
have summarized the relationship among the various algebras discussed in this section in Table~\ref{t1}.

\input{table1.tex}

\section{Lie Algebra of Gluonic Operators}
\label{s7}

In the last section we studied the actions of the operators of the
first three kinds, and obtained associative algebras closely related
to known algebras such as the Cuntz algebra.  It turns out that {\em products} of gluonic
operators with each other cannot be written as {\em finite} linear combinations of
these operators: they do not span an algebra under multiplication.  However
the {\em commutator} of two gluonic operators can be written as a finite
linear combination of gluonic operators. Thus gluonic operators acting on meson
states  form a Lie algebra, which we will call the {\em centrix}
algebra $\hatcentrix$.  This new Lie algebra we discover reduces to
the Virasoro algebra (up to an extension by \m{gl_{+\infty}}) in the
special case \m{\Lambda=1}.  Thus the centrix algebra is a sort of
generalization of the Lie algebra of vector fields on
a circle to non-commutative geometry. 

In a previous Letter \cite{leerajlett} we reported on the Lie algebra of
gluonic operators acting on glueball states. This is a similar but {\em
different} algebra, the cyclix algebra. The fact that the glueball
states are cyclically symmetric under permutations makes the
cyclix Lie algebra differ in an essential way from the centrix algebra. 

We will see that there is a proper ideal \m{{\salt}'\equiv \gl} 
of the Lie algebra $\hatcentrix$; hence, there is a quotient Lie algebra
$\centrix=\hatcentrix/\salt'$.  Furthermore, we will see that $\hatleftix$ and
$\hatrightix$ have some elements in common with $\hatcentrix$. This is perhaps
a bit surprising since they originally were introduced using quark
operators and  $\hatcentrix$ is the algebra of purely gluonic
operators acting on meson states. More
precisely, the generators \m{l^I_J,r^I_J} with \m{I,J} non-empty are in fact some
linear combinations of \m{\sigma^I_J}.  
Thus we can form quotients of $\hatcentrix$ by the algebras of $l^I_J$'s
and $r^I_J$'s.  This is the essentially new object we have discovered.

In this section, we will only focus on the action of $\s^I_J$ on 
a segment of the gluonic sequence of a single meson state.  We can capture the essence of 
Eq.(\ref{1.4}) by the following \footnote{Of course,
$\sigma^I_J$ acting on the gluonless state gives zero.}:
\begin{equation}
   \s^I_J s^{\dot{K}} = \sum_{\dot{K}_1 K_2 \dot{K}_3 = \dot{K}} \delta^{K_2}_{J} s^{\dot{K_1} I \dot{K_3}}.
\label{7.1}    
\end{equation}
This equation can be visualized in Fig.~\ref{f3}(d).  From Eq.(\ref{7.1}), we can see that the set of $\s^I_J$'s 
with all possible non-empty sequences $I$ and $J$'s is linearly independent. 

\begin{figure}
\epsfxsize=5in
\centerline{\epsfbox{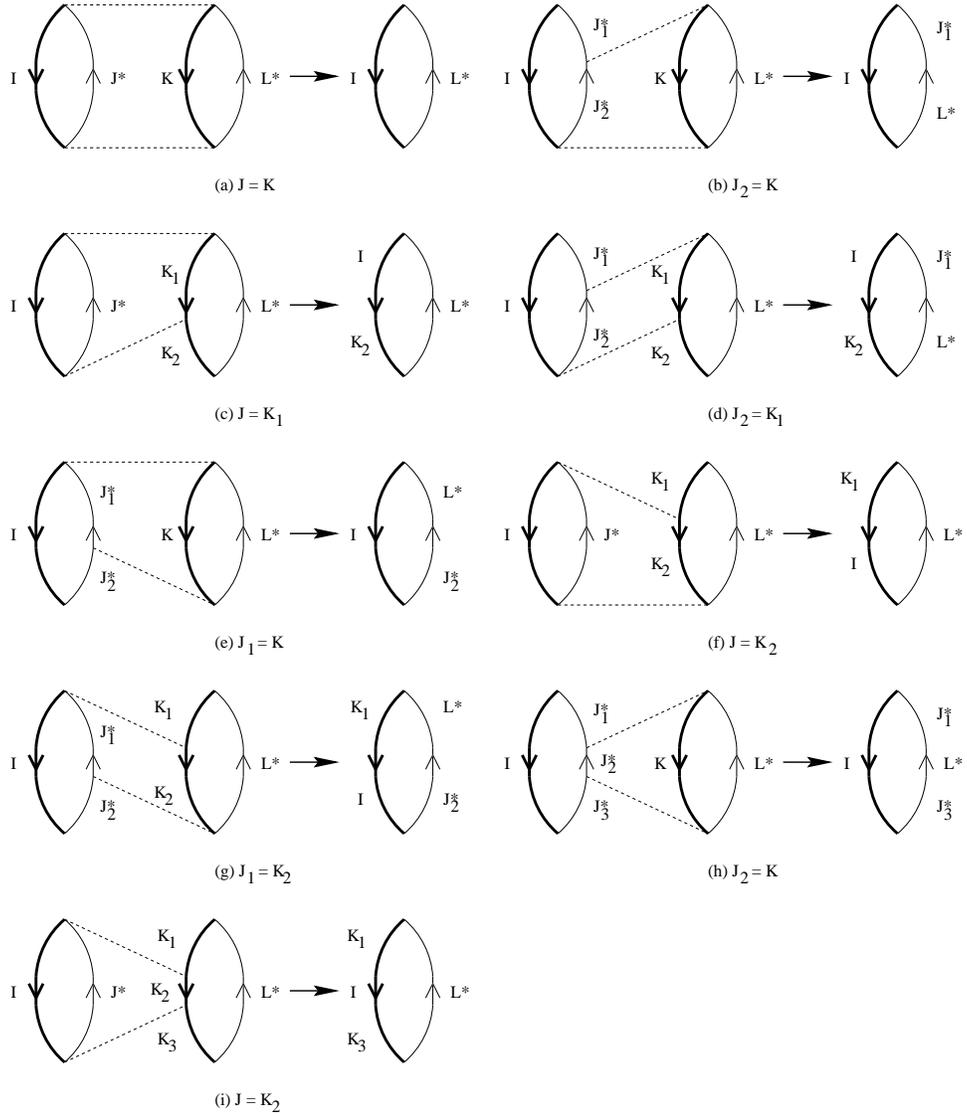}}
\caption{\em  Diagrammatic representations of the first 9 terms on the R.H.S.
of 
Eq.(\ref{7.4}).}
\label{f5.3}
\end{figure}

Unlike the case of the algebra $\hatleftix$ considered in Section~\ref{s3}, we cannot define the 
composite operator $\s^I_J \s^K_L$ by the requirement that its action on $\s^{\dot{M}}$ give 
$\s^I_J (\s^K_L s^{\dot{M}})$ because in general this action cannot be written as a linear 
combination of the $\s$'s \cite{lee}.  Nonetheless, the Lie bracket between two $\s$'s is well-defined by the 
requirement that
\begin{equation}
   \left( \lbrack \s^I_J, \s^K_L\rbrack \right) s^{\dot{P}} \equiv \s^I_J \left( \s^K_L s^{\dot{P}} \right) -
   \s^K_L \left( \s^I_J s^{\dot{P}} \right).
\label{7.3}
\end{equation}
for any arbitrary sequence $\dot{P}$.  Then it can be shown \cite{lee}
(by a tedious but straightforward calculation) that the expression of the Lie bracket is
\begin{eqnarray}
   \lefteqn{ \left[ \s^I_J, \s^K_L \right] =
   \delta^K_J \s^I_L + \sum_{J_1 J_2 = J} \delta^K_{J_2} 
   \s^I_{J_1 L} + \sum_{K_1 K_2 = K} \delta^{K_1}_J \s^{I K_2}_L } \nonumber \\
   & & + \sum_{\begin{array}{l}
		  J_1 J_2 = J \\
		  K_1 K_2 = K
	       \end{array}}
   \delta^{K_1}_{J_2} \s^{I K_2}_{J_1 L} 
   + \sum_{J_1 J_2 = J} \delta^K_{J_1} \s^I_{L J_2}
   + \sum_{K_1 K_2 = J} \delta^{K_2}_J \s^{K_1 I}_L \nonumber \\
   & & + \sum_{\begin{array}{l}
		  J_1 J_2 = J \\
		  K_1 K_2 = K
	       \end{array}}
   \delta^{K_2}_{J_1} \s^{K_1 I}_{L J_2}
   + \sum_{J_1 J_2 J_3 = J} \delta^K_{J_2} \s^I_{J_1 L J_3} 
   + \sum_{K_1 K_2 K_3 = K} \delta^{K_2}_J \s^{K_1 I K_3}_L \nonumber \\
   & & - (I \leftrightarrow K, J \leftrightarrow L). 
\label{7.4}
\end{eqnarray}
The diagrammatic representations of the first 9 terms are given in Fig.~\ref{f5.3}.  We will call the Lie 
algebra defined by Eq.(\ref{7.4}) the {\em centrix algebra} $\hat{\Sigma}_{{\Lambda}}$.

Now let us note the following identities that follow easily from the action of $\sigma^I_J$:
\[ \left( \s^I_J - \sum_{i=1}^{\Lambda} \s^{iI}_{iJ} \right) s^{\dot{K}} 
   = \sum_{K_1 \dot{K}_2 = \dot{K}} \delta^{K_1}_J s^{I\dot{K}_2} \]
and
\[ \left( \s^I_J - \sum_{j=1}^{\Lambda} \s^{Ij}_{Jj} \right) s^{\dot{K}} 
   = \sum_{\dot{K}_1 K_2 = \dot{K}} \delta^{K_2}_J s^{\dot{K}_1 I}. \]
These are exactly the action of the opeartors \m{l^I_J} and \m{r^I_J}
we defined previously. Thus we have,
\begin{eqnarray}
   l^I_J & = & \s^I_J - \sum_{i=1}^{{\Lambda}} \s^{iI}_{iJ}\mbox{; and}
\label{7.6} \\
   r^I_J & = & \s^I_J - \sum_{j=1}^{{\Lambda}} \s^{Ij}_{Jj}.
\label{7.6.1}
\end{eqnarray}
Of course we must require that \m{I} and \m{J} be non-empty for this
to be true. 

The operators \m{l^I_J}'s with non-empty sequences \m{I} and \m{J} form
a subalgebra \m{\hatleftix'} of $\hatleftix$, the operators $r^I_J$'s form another subalgebra
$\hatrightix'$ of $\hatrightix$, and the operators $l^I_J$'s and $r^I_J$'s together form yet another subalgebra
$\hatmultix'$ of $\hatmultix$.  What we have just seen is
that $\hatleftix'$, $\hatrightix'$ and $\hatmultix'$ can also be viewed as subalgebras of \m{\hatcentrix}; 
in fact they are even ideals of \m{\hatcentrix}.  This will be
evident from the (more general) commutation relations between
\m{\sigma^I_J} and \m{l^{\dot{I}}_{\dot{J}}}, etc. given below.

Analogously, we can define a vector space ${\salt'}$ spanned by $f^I_J$
with non-empty $I$ and $J$. By the same counting argument as before, this
is also isomorphic to \m{\gl}. We have already seen that :
\begin{equation}
   f^K_L = l^K_L - \sum_{j=1}^{\Lambda} l^{Kj}_{Lj}
\label{7.7}
\end{equation}       
and 
\begin{equation}
   f^K_L = r^K_L - \sum_{i=1}^{\Lambda} r^{iK}_{iL}.
\end{equation}
Thus we have
\beq
  f^I_J = \sigma^I_J - \sum_{j=1}^{\Lambda} \sigma^{Ij}_{Jj} - \sum_{i=1}^{\Lambda} \sigma^{iI}_{iJ}
  + \sum_{i,j=1}^{\Lambda} \sigma^{iIj}_{iJj}.
\label{7.11}
\eeq
Thus \m{\salt'} is an ideal of \m{\hatcentrix} as well. 

\begin{figure}[ht]
\epsfxsize=5in
\centerline{\epsfbox{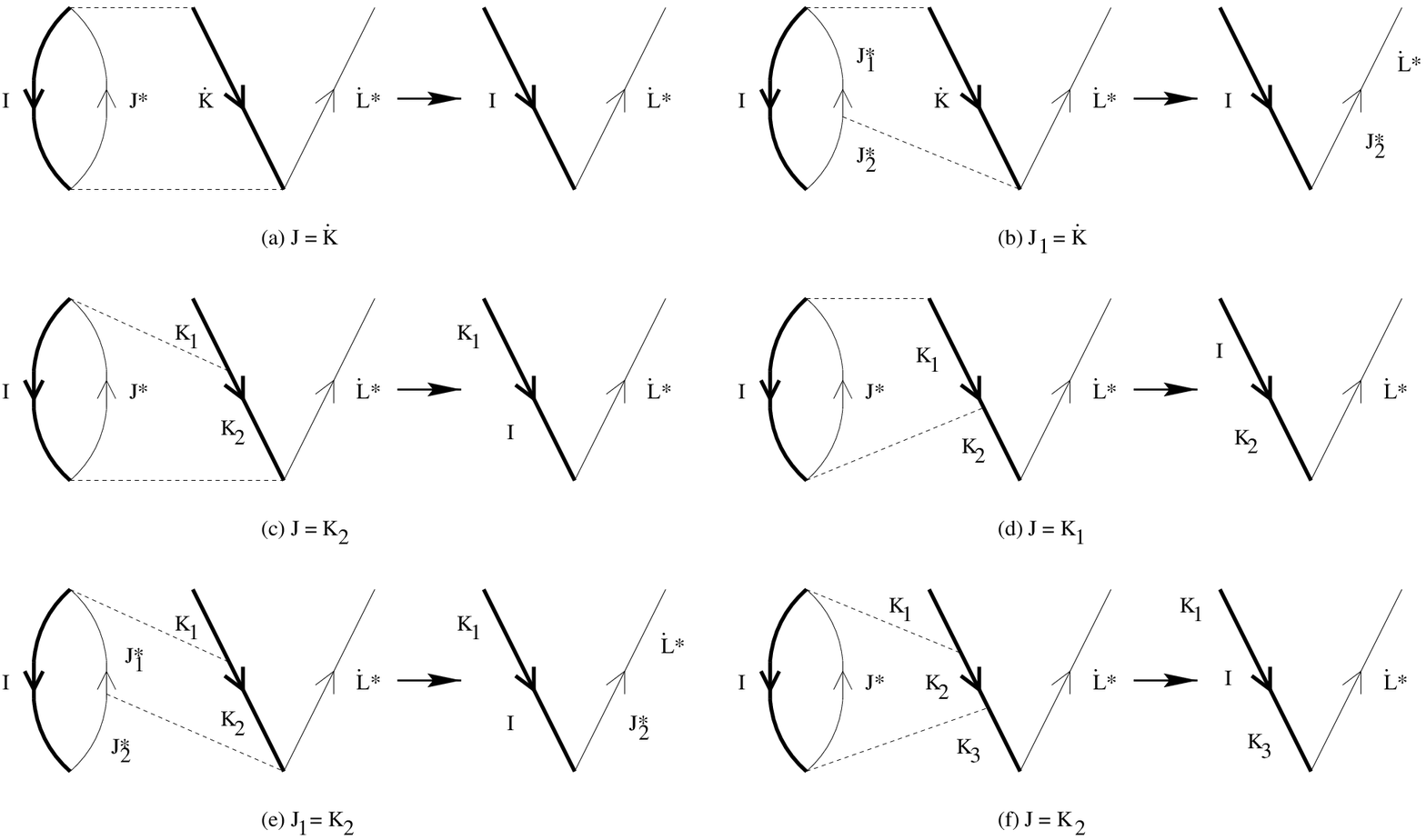}}
\caption{\em  Diagrammatic representation of Eq.(\ref{7.8}).  Only the first 6 terms on the 
R.H.S. of Eq.(\ref{7.8}) are shown here.}
\label{f7.2}
\end{figure}

\begin{figure}
\epsfxsize=5in
\centerline{\epsfbox{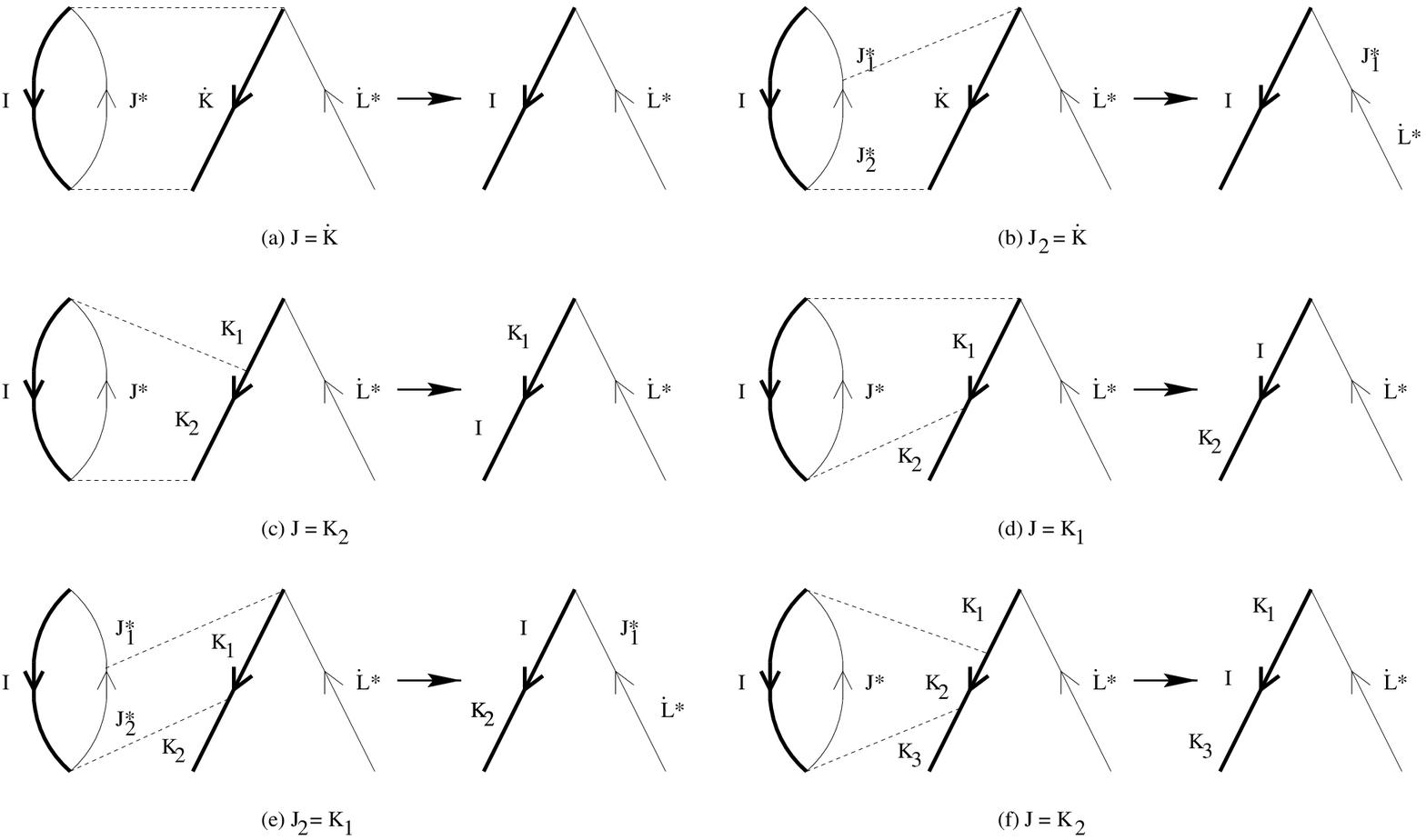}}
\caption{\em  Diagrammatic representation of Eq.(\ref{7.8.1}).  Only the first 6 terms on 
the R.H.S. of Eq.(\ref{7.8.1}) are shown here.}
\label{f7.2.1}
\end{figure}

\begin{figure}[ht]
\epsfxsize=2.5in
\centerline{\epsfbox{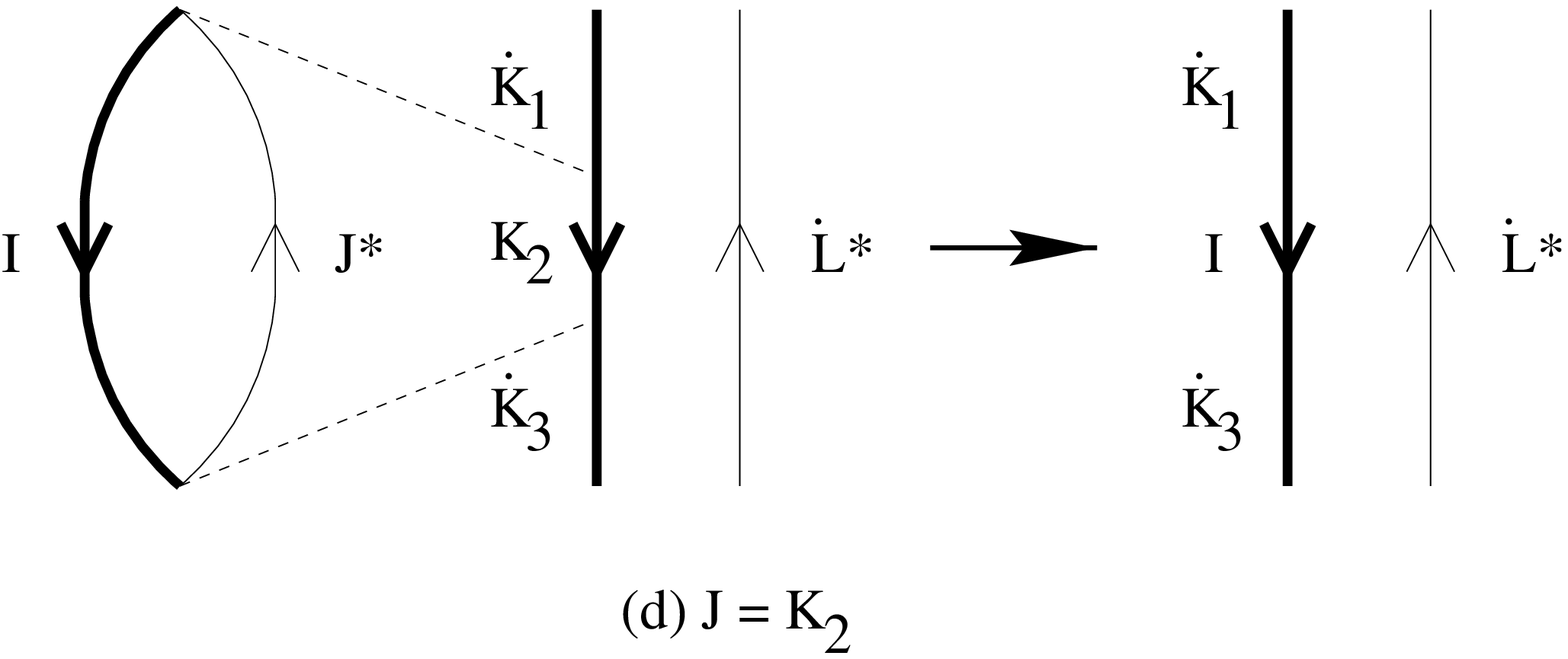}}
\caption{\em  Diagrammatic representation of Eq.(\ref{7.9}).  Only the first term on the 
R.H.S. of Eq.(\ref{7.9}) is shown here.}
\label{f7.3}
\end{figure}

Now we give the commutation relations between the centrix and multix operators:
\begin{eqnarray}
   \lefteqn{\left[ \s^I_J, l^{\dot{K}}_{\dot{L}} \right] = \delta^{\dot{K}}_J l^I_{\dot{L}} + 
   \sum_{J_1 J_2 = J} \delta^{\dot{K}}_{J_1} l^I_{\dot{L} J_2} +
   \sum_{K_1 K_2 = \dot{K}} \delta^{K_2}_J l^{K_1 I}_{\dot{L}} } \nonumber \\
   & & + \sum_{K_1 K_2 = \dot{K}} \delta^{K_1}_J l^{I K_2}_{\dot{L}} +
   \sum_{\begin{array}{l}
	    J_1 J_2 = J \\
	    K_1 K_2 = \dot{K}
	 \end{array}}
   \delta^{K_2}_{J_1} l^{K_1 I}_{\dot{L} J_2}
   + \sum_{K_1 K_2 K_3 = \dot{K}} \delta^{K_2}_J l^{K_1 I K_3}_{\dot{L}} \nonumber \\
   & & - \delta^I_{\dot{L}} l^{\dot{K}}_J -
   \sum_{I_1 I_2 = I} \delta^{I_1}_{\dot{L}} l^{\dot{K} I_2}_J -
   \sum_{L_1 L_2 = \dot{L}} \delta^I_{L_2} l^{\dot{K}}_{L_1 J} \nonumber \\
   & & - \sum_{L_1 L_2 = \dot{L}} \delta^I_{L_1} l^{\dot{K}}_{J L_2} -
   \sum_{\begin{array}{l}
   	    L_1 L_2 = \dot{L} \\
   	    I_1 I_2 = I
   	 \end{array}}
   \delta^{I_1}_{L_2} l^{\dot{K} I_2}_{L_1 J} 
   - \sum_{L_1 L_2 L_3 = \dot{L}} \delta^I_{L_2} l^{\dot{K}}_{L_1 J L_3};
\label{7.8} \\
   \lefteqn{\left[ \s^I_J, r^{\dot{K}}_{\dot{L}} \right] = \delta^{\dot{K}}_J r^I_{\dot{L}} + 
   \sum_{J_1 J_2 = J} \delta^{\dot{K}}_{J_2} r^I_{J_1 \dot{L}} +
   \sum_{K_1 K_2 = \dot{K}} \delta^{K_2}_J r^{K_1 I}_{\dot{L}} } \nonumber \\
   & & + \sum_{K_1 K_2 = \dot{K}} \delta^{K_1}_J r^{I K_2}_{\dot{L}} +
   \sum_{\begin{array}{l}
	    J_1 J_2 = J \\
	    K_1 K_2 = \dot{K}
	 \end{array}}
   \delta^{K_1}_{J_2} r^{I K_2}_{J_1 \dot{L}}
   + \sum_{K_1 K_2 K_3 = \dot{K}} \delta^{K_2}_J r^{K_1 I K_3}_{\dot{L}} \nonumber \\
   & & - \delta^I_{\dot{L}} r^{\dot{K}}_J -
   \sum_{I_1 I_2 = I} \delta^{I_2}_{\dot{L}} r^{I_1 \dot{K}}_J -
   \sum_{L_1 L_2 = \dot{L}} \delta^I_{L_2} r^{\dot{K}}_{L_1 J} \nonumber \\
   & & - \sum_{L_1 L_2 = \dot{L}} \delta^I_{L_1} r^{\dot{K}}_{J L_2} -
   \sum_{\begin{array}{l}
   	    L_1 L_2 = \dot{L} \\
   	    I_1 I_2 = I
   	 \end{array}}
   \delta^{I_2}_{L_1} r^{I_1 \dot{K}}_{J L_2} 
   - \sum_{L_1 L_2 L_3 = \dot{L}} \delta^I_{L_2} r^{\dot{K}}_{L_1 J L_3} \mbox{; and}
\label{7.8.1} \\
   \lefteqn{\left[ \s^I_J, f^{\dot{K}}_{\dot{L}} \right] = 
   \sum_{\dot{K}_1 K_2 \dot{K}_3 = \dot{K}} \delta^{K_2}_J f^{\dot{K}_1 I \dot{K}_3}_{\dot{L}}
   - \sum_{\dot{L}_1 L_2 \dot{L}_3 = \dot{L}} \delta^I_{L_2} f^{\dot{K}}_{\dot{L}_1 J \dot{L}_3}. }
\label{7.9}
\end{eqnarray}
Eqs.(\ref{7.8}), (\ref{7.8.1}) and (\ref{7.9}) are illustrated in 
Figs.~\ref{f7.2}, \ref{f7.2.1} and \ref{f7.3} respectively.

These prove, in particular, that the algebras $\hatleftix'$, $\hatrightix'$ and $\salt'$ are proper ideals of the 
centrix algebra.  In addition, Eqs.(\ref{7.8}) and (\ref{7.8.1}) show that an element which does not 
belong to $\hatleftix'$ in the centrix algebra is an outer derivation of the algebra $\hatleftix$, i.e., the Lie 
algebra associated with the extended Cuntz algebra, and an element which does not belong to $\hatrightix'$ in the 
centrix algebra is an outer derivation of the algebra $\hatrightix$.  Let us summarize the
various relationships among the Lie algebras in Table~\ref{t1_1}.

\input{table1_1.tex}

It follows from the above discussion that there are various ways to form different quotient algebras from the 
centrix algebra.  For instance, we can make the set of all cosets $\s^I_J + \salt'$ into a quotient algebra 
$\centrix$.  Furthermore, within this quotient algebra, all cosets of the form $l^I_J + \salt$ span a 
proper ideal $\leftix'$, and all cosets of the form $r^I_J + \salt$ span another ideal 
$\rightix'$.  Thus we can extract the quotient Lie algebra
\beq
	\vectrix \equiv \hatcentrix / \hatmultix' = \centrix / \multix'.
\eeq

In the simplest case \m{\Lambda=1}, $\vectrix = {\cal V}_1$ is just the algebra of vector fields on a circle.  
Indeed, now all integer sequences are just repetitions of the number 1 a number of
times. Write  $s^{\dot{K}}$ as $s^{\#(\dot{K})}$, $l^I_J$ as $l^{\#(I)}_{\#(J)}$, and $\s^I_J$ as 
$\s^{\#(I)}_{\#(J)}$.  (Note that $r^{\#(I)}_{\#(J)} = l^{\#(I)}_{\#(J)}$.)  It can be easily seen that $M'_1$ is 
spanned by the cosets of the forms $l^a_1 + F'_1$ and $l^1_b + F'_1$, and $\Sigma_1$ is spanned by the cosets 
$l^a_1 + \saltone$, $l^1_b + \saltone$, $\s^a_1 + \saltone$ and $\s^1_b + \saltone$, where $a$ and $b$ run over all
positive integers.  It is a straightforward matter to verify the following Lie brackets, {\em modulo} \m{\saltone}:
\begin{eqnarray}
   \left[ l^a_1 , l^c_1  \right] & = & 0=\left[ l^a_1, l^1_d\right]=\left[ l^1_b , l^1_d\right] ; \nonumber \\
   \left[ \s^a_1, \s^c_1  \right] & = & (c - a)\s^{a+c-1}  ; \nonumber \\
   \left[ \s^a_1, \s^1_d \right] & = &
   \left\{ \begin{array}{ll}
   (2 - a - d) \s^{a-d+1}_1 & \mbox{if $d \leq a$, or} \\
   (2 - a - d) \s^1_{d-a+1} & \mbox{if $a \leq d$;}
   \end{array} \right. \nonumber \\
   \left[ \s^1_b, \s^1_d  \right] & = & (b - d) \s^1_{b+d-1}; \nonumber \\
   \left[ \s^a_1 , l^c_1  \right] & = & (c - 1) l^{a+c-1}_1; \nonumber \\
   \left[ \s^1_b , l^c_1  \right] & = & \left\{
   \begin{array}{ll}
      (c - 1) l^{c-b+1}_1 & \mbox{if $b \leq c$, or} \\
      (c - 1) l^1_{b-c+1} & \mbox{if $c \leq b$;}
   \end{array} \right. \nonumber \\
   \left[ \s^1_b , l^1_d\right] & = & 
   - (d - 1) \s^1_{b+d-1} \mbox{; and} \nonumber \\
   \left[ \s^a_1 , l^1_d \right] & = & \left\{
   \begin{array}{ll}
      - (d - 1) l^1_{d-a+1} & \mbox{if $a \leq d$, or} \\
      - (d - 1) l^{a-d+1}_1 & \mbox{if $d \leq a$.}
   \end{array} \right. .
\label{8.1}
\end{eqnarray}

Let us make the following identifications:
\begin{eqnarray}
   l^a_1  & \rightarrow & z^{a-1} ; \nonumber \\
   l^1_b  & \rightarrow & z^{-(b-1)} ; \nonumber \\
   \s^a_1  & \rightarrow & z^{a-1} \frac{d}{dz} = - L_{a-1} 
   \mbox{; and} \nonumber \\
   \s^1_b  & \rightarrow & z^{-(b-1)} \frac{d}{dz} = - L_{1-b}.
\label{8.2}
\end{eqnarray}
Here $z$ is a complex number with $|z| = 1$.  Note that the $L$ here is 
{\em not} an integer sequence.  Then Eq.(\ref{8.1}) becomes
\begin{eqnarray}
   \left[ z^p, z^q \right] & = & 0 ; \nonumber \\
   \left[ L_p, L_q \right] & = & (p - q) L_{p+q} \mbox{; and} \nonumber \\
   \left[ L_p, z^q \right] & = & -q z^{p+q-1},
\end{eqnarray}    
where $p$ and $q$ are integers which may be positive, negative or zero.
Thus this is the extension of the Virasoro algebra without any central element by the 
functions on a unit circle on the complex plane.  We also see that the
algebra \m{{\cal V}_1} is just the Lie algebra of vector fields on the
circle, spanned by all \m{L_p}'s for \m{p} being all integers.

More generally, the generators of \m{\hatcentrix} act as derivations on the
Cuntz algebra. When we quotient out $\hatcentrix$ by the ideal $\hatmultix'$, we are
basically extracting the part which corresponds to `inner
derivations' (up to some extensions by standard algebras such as $\gl$.)
Thus the vectrix algebra \m{\vectrix } should be viewed as a generalization of the
algebra of vector fields of a circle. It is the  Lie algebra of
vector fields on a sort of non-commutative generalization of the
circle, which has the Cuntz algebra as its algebra of functions. It
would be interesting  to investigate this further.

\section{Grand Algebra Acting on Open Strings}
\label{s6}

Now we piece together the algebras of current and gluonic operators to 
form a grand Lie algebra. In general the Hamiltonian of a matrix model in
the large-\m{N} limit will be an element of this algebra. 
We are going to write down the Lie brackets of various operators, 
and we will see that this grand Lie algebra is a sum, though not a direct sum,
of four subalgebras, and there are two proper ideals in this grand Lie algebra.

Let us begin by considering operators of the first kind.  It is
easy to see that all $\bar{\Xi}^{\lambda_1}_{\lambda_2} \otimes f^{\dot{I}}_{\dot{J}}
\otimes \Xi^{\lambda_3}_{\lambda_4}$'s span a Lie algebra
\m{\grandsalt} which is isomoprphic to $gl(\bar{\Lambda}_F) \otimes
{\mathit F}_{\Lambda} \otimes gl(\Lambda_F)$. Of course this in turn is
just isomorphic to \m{gl_{+\infty}}, by an appropriate one-to-one
correspondence with natural numbers.

Next, let us move to operators of the second kind.  We deduce from Eqs.(\ref{2.0.1}) and 
(\ref{2.2}) that the Lie bracket between two operators of this form is
\begin{eqnarray}
   \lefteqn{ \left[ \bar{\Xi}^{\lambda_1}_{\lambda_2} \otimes l^{\dot{I}}_{\dot{J}} \otimes 1,
   \bar{\Xi}^{\lambda_3}_{\lambda_4} \otimes l^{\dot{K}}_{\dot{L}} \otimes 1 \right] = } \nonumber \\
   & & \delta^{\lambda_3}_{\lambda_2} \bar{\Xi}^{\lambda_1}_{\lambda_4} \otimes
   \left( \delta^{\dot{K}}_{\dot{J}} l^{\dot{I}}_{\dot{L}} + 
   \sum_{\dot{J}_1 J_2 = \dot{J}} \delta^{\dot{K}}_{\dot{J}_1} l^{\dot{I}}_{\dot{L} J_2} + 
   \sum_{\dot{K}_1 K_2 = \dot{K}} \delta^{\dot{K}_1}_{\dot{J}} l^{\dot{I} K_2}_{\dot{L}} \right) \otimes 1 
   \nonumber \\
   & & - \delta^{\lambda_1}_{\lambda_4} \bar{\Xi}^{\lambda_3}_{\lambda_2} \otimes
   \left( \delta^{\dot{I}}_{\dot{L}} l^{\dot{K}}_{\dot{J}} + 
   \sum_{\dot{L}_1 L_2 = \dot{L}} \delta^{\dot{I}}_{\dot{L}_1} l^{\dot{K}}_{\dot{J} L_2} + 
   \sum_{\dot{I}_1 I_2 = \dot{I}} \delta^{\dot{I}_1}_{\dot{L}} l^{\dot{K} I_2}_{\dot{J}} \right) \otimes 1
\label{6.1}
\end{eqnarray}
Thus operators of the second kind form another subalgebra $\grandhatL$of the grand algebra. 

The Lie bracket between two operators of the third kind can be similarly derived from Eqs.(\ref{2.0.2}) and 
(\ref{3.9}):
\begin{eqnarray}
   \lefteqn{ \left[ 1 \otimes r^{\dot{I}}_{\dot{J}} \otimes \Xi^{\lambda_1}_{\lambda_2},
   1 \otimes r^{\dot{K}}_{\dot{L}} \otimes \Xi^{\lambda_3}_{\lambda_4} \right] = } \nonumber \\
   & & \delta^{\lambda_3}_{\lambda_2} 1 \otimes \left( \delta^{\dot{K}}_{\dot{J}} r^{\dot{I}}_{\dot{L}} +
   \sum_{J_1 \dot{J}_2 = \dot{J}} \delta^{\dot{K}}_{\dot{J}_2} r^{\dot{I}}_{J_1 \dot{L}} + 
   \sum_{K_1 \dot{K}_2 = \dot{K}} \delta^{\dot{K}_2}_{\dot{J}} r^{K_1 \dot{I}}_{\dot{L}} \right) \otimes 
   \Xi^{\lambda_1}_{\lambda_4} \nonumber \\
   & & - \delta^{\lambda_1}_{\lambda_4} 1 \otimes \left( \delta^{\dot{I}}_{\dot{L}} r^{\dot{K}}_{\dot{J}} +
   \sum_{L_1 \dot{L}_2 = \dot{L}} \delta^{\dot{I}}_{\dot{L}_2} r^{\dot{K}}_{L_1 \dot{J}} + 
   \sum_{I_1 \dot{I}_2 = \dot{I}} \delta^{\dot{I}_2}_{\dot{L}} r^{I_1 \dot{K}}_{\dot{J}} \right) \otimes 
   \Xi^{\lambda_3}_{\lambda_2}. 
\label{6.2}
\end{eqnarray}
This equation shows clearly that all operators of the third kind form yet another subalgebra of the 
grand algebra.

Let us consider the Lie bracket relations between operators of different kinds.
We can derive the Lie bracket relations 
between an operator of the first kind, and either an operator of the second 
kind or third kind with the help of Eqs.(\ref{3.5}) and (\ref{3.8}) as follows:
\begin{eqnarray}
   \lefteqn{ \left[ \bar{\Xi}^{\lambda_1}_{\lambda_2} \otimes l^{\dot{I}}_{\dot{J}} \otimes 1,
   \bar{\Xi}^{\lambda_3}_{\lambda_4} \otimes f^{\dot{K}}_{\dot{L}} \otimes \Xi^{\lambda_5}_{\lambda_6} \right] = } 
   \nonumber \\
   & & \delta^{\lambda_3}_{\lambda_2} \bar{\Xi}^{\lambda_1}_{\lambda_4} \otimes
   \sum_{\dot{K}_1 \dot{K}_2 = \dot{K}} \delta^{\dot{K}_1}_{\dot{J}} f^{\dot{I} \dot{K}_2}_{\dot{L}}
   \otimes \Xi^{\lambda_5}_{\lambda_6}
   - \delta^{\lambda_1}_{\lambda_4} \bar{\Xi}^{\lambda_3}_{\lambda_2} \otimes
   \sum_{\dot{L}_1 \dot{L}_2 = \dot{L}} \delta^{\dot{I}}_{\dot{L}_1} f^{\dot{K}}_{\dot{J} \dot{L}_2}
   \otimes \Xi^{\lambda_5}_{\lambda_6}
\end{eqnarray}
and
\begin{eqnarray}
   \lefteqn{ \left[ 1 \otimes r^{\dot{I}}_{\dot{J}} \otimes \Xi^{\lambda_1}_{\lambda_2},
   \bar{\Xi}^{\lambda_3}_{\lambda_4} \otimes f^{\dot{K}}_{\dot{L}} \otimes \Xi^{\lambda_5}_{\lambda_6} \right] = } 
   \nonumber \\
   & & \delta^{\lambda_5}_{\lambda_2} \bar{\Xi}^{\lambda_3}_{\lambda_4} \otimes
   \sum_{\dot{K}_1 \dot{K}_2 = \dot{K}} \delta^{\dot{K}_2}_{\dot{J}} f^{\dot{K}_1 \dot{I}}_{\dot{L}}
   \otimes \Xi^{\lambda_1}_{\lambda_6} 
   - \delta^{\lambda_1}_{\lambda_6} \bar{\Xi}^{\lambda_3}_{\lambda_4} \otimes
   \sum_{\dot{L}_1 \dot{L}_2 = \dot{L}} \delta^{\dot{I}}_{\dot{L}_2} f^{\dot{K}}_{\dot{L}_1 \dot{J}}
   \otimes \Xi^{\lambda_5}_{\lambda_2}.
\end{eqnarray}
These equations show that the subalgebra of the operators of the first kind form a proper 
ideal of the Lie algebra spanned by the operators of the first, second and third kinds.
The Lie bracket between a typical operator of the second and that of the third kind is
\begin{equation}
   \left[ \bar{\Xi}^{\lambda_1}_{\lambda_2} \otimes l^{\dot{I}}_{\dot{J}} \otimes 1,
   1 \otimes r^{\dot{K}}_{\dot{L}} \otimes \Xi^{\lambda_3}_{\lambda_4} \right] =
   \bar{\Xi}^{\lambda_1}_{\lambda_2} \otimes \left[ l^{\dot{I}}_{\dot{J}}, r^{\dot{K}}_{\dot{L}} \right] \otimes
   \Xi^{\lambda_3}_{\lambda_4},
\end{equation}
where $\left[ l^{\dot{I}}_{\dot{J}}, r^{\dot{K}}_{\dot{L}} \right]$ is given by the R.H.S. of Eq.(\ref{7.8.2}).  
As a result, this Lie bracket gives a linear combination of operators of the first kind.  

Finally, consider operators of the fourth kind.  The Lie bracket relations between an operator of the fourth kind, 
and an operator of any kind, are as follows:
\begin{eqnarray}
   \left[ 1 \otimes \s^I_J \otimes 1, 1 \otimes \s^K_L \otimes 1 \right] & = &
   1 \otimes \left[ \s^I_J, \s^K_L \right] \otimes 1; \\
   \left[ 1 \otimes \s^I_J \otimes 1, 1 \otimes r^{\dot{K}}_{\dot{L}} \otimes \Xi^{\lambda_1}_{\lambda_2}
   \right] & = & 1 \otimes \left[ \s^I_J, r^{\dot{K}}_{\dot{L}} \right] 
   \otimes \Xi^{\lambda_1}_{\lambda_2}; \\
   \left[ 1 \otimes \s^I_J \otimes 1, \bar{\Xi}^{\lambda_1}_{\lambda_2} \otimes l^{\dot{K}}_{\dot{L}} \otimes 1
   \right] & = & \Xi^{\lambda_1}_{\lambda_2} \otimes \left[ \s^I_J, l^{\dot{K}}_{\dot{L}} \right]
   \otimes 1 \mbox{; and} \\
   \left[ 1 \otimes \s^I_J \otimes 1, \bar{\Xi}^{\lambda_1}_{\lambda_2} \otimes f^{\dot{K}}_{\dot{L}}
   \otimes \Xi^{\lambda_3}_{\lambda_4} \right] & = & \bar{\Xi}^{\lambda_1}_{\lambda_2}
   \otimes \left[ \s^I_J, f^{\dot{K}}_{\dot{L}} \right] \otimes \Xi^{\lambda_3}_{\lambda_4}.
\end{eqnarray}
From the above four Lie brackets and Eqs.(\ref{7.4}), (\ref{7.8}), (\ref{7.8.1}) and 
(\ref{7.9}), we conclude that all operators of the fourth kind form a subalgebra of the 
grand algebra, and that the Lie bracket between an operator of the fourth kind and an 
operator of the third, second or first kind produces an operator of the third, second or 
first kind respectively.  Operators of these four kinds span the whole grand algebra of 
operators acting on mesons or open strings.  Operators of the first, second and third kinds 
together span a proper ideal of the grand algebra.  We have summarized these results in Table~\ref{t2}.

\input{table2.tex}

\section{Spin Chain Models}
\label{s9}   

Now that we have understood the algebraic properties of current and gluonic operators and 
their actions on open string states, let us give some physical models in which these 
operators and states show up.  We will discuss two families of models in detail: 
{\em multi}-matrix models {\em integrable} in the large-$N$ limit associated with 
integrable quantum spin chain models with open boundary conditions in this section, 
and quantum chromodynamics in the large-$N$ limit in the next section.  

In a previous Letter \cite{leerajlett}, we described the procedure for transcribing quantum spin chain
models satisfying the periodic boundary condition into the corresponding multi-matrix models.  The same procedure 
can be used to establish a one-to-one corespondence between quantum spin chain models satisfying open boundary
conditions and multi-matrix models.  The major modification is that boundary terms in the Hamiltonian of a spin
chain corresponds to elements in $\hatleftix$ and $\hatrightix$.  Suppose a Hamiltonian of a matrix model
is a linear combination $H = \sum_{IJ} (h^J_I \s^I_J + a^J_I l^I_J + b^J_I r^I_J)$ where
$h^J_I$, $a^J_I$ and $b^J_I$ are non-zero coefficients only if $I$ and $J$ have the same
number of indices.  Such linear combinations form a subalgebra of $\hatcentrix$; 
let us call it $\hat{\Sigma}^0_{\Lambda}$.  Then {\em there is an isomorphism between 
multi-matrix models whose Hamiltonians are in $\hat{\Sigma}^0_{\Lambda}$ and quantum
spin chains with interactions involving neighborhoods of spins, and with open
boundary conditions}.  Furthermore, if the quantum spin chain model is exactly solvable, so
is the associated multi-matrix model, and vice versa. 

These matrix models preserve the number of `gluons' or string bits. Such models can be used to study interesting 
string phenomena whenever the string tension becomes so large that the ground state favors as many string bits as
possible \cite{KlSu}.  Since matrix models are appearing in many different contexts, the ideas in Ref.\cite{KlSu} 
are probably intersting even outside string theory. 

Our work also suggests
a deep relationship between integrable spin chain models and our Lie
algebras: to each solution of the Yang-Baxter-Sklyanin relations,
there is a maximal abelian subalgebra of the centrix algebra. This is
just the subalgebra spanned by the conserved quantities of the
corresponding spin chain: each conserved quantity is an element of our
centrix algebra. Thus it emerges  that the Lie algebras we have found
are the underlying symmetries of many integrable models. We intend to
develop these ideas in the future into a systematic theory. For now
we content ourselves with a list of integrable multi-matrix models.

\begin{itemize}
\item {\em the six-vertex model and the corresponding integrable open quantum spin-$\frac{1}{2}$ chain model}
\cite{Sk, GoRuSi, AlBaBa}.  A typical quantum state $\Phi$ can be characterized 
by a sequence of $c$ 2-dimensional column vectors, where $c$ is the number of sites in the 
open chain.  Spin-up and spin-down states at the $j$-th site are characterized by the 
$j$-th column vectors being
\[ \left( \begin{array}{c} 1 \\ 0 \end{array} \right) \; \mbox{and} \;
   \left( \begin{array}{c} 0 \\ 1 \end{array} \right) \]
respectively.  The Hamiltonian $H^{\spin}_{\rm XXZ}$ of this spin chain model is
\begin{eqnarray}
   H^{\spin}_{\rm XXZ} & = & \frac{1}{2 \sin \gamma} \left[ \sum_{j=1}^{c-1} 
   (\tau_j^x \tau_{j+1}^x + \tau_j^y \tau_{j+1}^y + \cos \gamma \tau_j^z
\tau_{j+1}^z)
   \right. \nonumber \\
   & & + \left. i \sin \gamma ( \coth \xi_- \tau_1^z + \coth \xi_+ \tau_c^z )
\right]
\label{9.1}
\end{eqnarray}
where $\gamma \in (0, \pi)$ and both $\xi_-$ and $\xi_+$ are arbitrary constants.  In 
Eq.(\ref{9.1}), $\tau_j^x$, $\tau_j^y$ and $\tau_j^z$ are Pauli matrices at site $j$.  They 
act on the $j$-th column vector.  Two Pauli matrices at different sites (i.e., with 
different subscripts) are commuting.  This is actually an integrable XXZ model with an open 
boundary condition.

We can paraphrase the model in terms of the states and operators of the matrix model 
introduced in Section~\ref{s2} as follows.  A single meson state corresponds to an open spin 
chain.  We allow the existence of only one possible quantum state other than color for both 
quarks and antiquarks.  Thus the notations for quarks and antiquarks in the expressions for 
single meson states can be suppressed.  Each gluon corresponds to a spin.  There are 2 
possible quantum states (i.e., ${\Lambda} = 2$) other than color for a gluon.  $a^{\dagger}(1)$ 
(color indices are suppressed) corresponds to a spin-up state, whereas $a^{\dagger}(2)$ 
corresponds to a spin-down state.  Therefore a typical quantum state of a spin chain can be 
denoted by $s^K$, where $K$ is an integer sequence of $c$ numbers, each of which is either 1 or 
2.

The bulk term of the Hamiltonian $H^{\matrix}_{\rm XXZ}$ of the corresponding integrable matrix model in the
large-$N$ limit can be determined by the almost same procedure outlined in Ref.\cite{leerajlett}, except a minor
change in the ranges of summations in some formulae there.  The boundary terms on the left can be obtained from the
correspondences $\tau^z_1 \leftrightarrow l^1_1 - l^2_2$, $\tau^x_1 + {\rm i} \tau^y_1 \leftrightarrow 2 l^1_2$ and 
$\tau^x_1 - {\rm i} \tau^y_1 \leftrightarrow 2 l^2_1$.  The boundary terms on the right can be determined 
analogously.  Consequently,
\begin{eqnarray}
   H^{\matrix}_{\rm XXZ} & = & \frac{1}{2 \sin \gamma} \left\{ 2 ( \s_{12}^{21} + 		
   \s_{21}^{12} + \cos\gamma (\s_{11}^{11} - \s_{12}^{12} - \s_{21}^{21} + \s_{22}^{22} )
   \right. \nonumber \\
   & & + i \left. \sin\gamma \lbrack \coth\xi_- (l_1^1 - l_2^2) + \coth\xi_+ (r_1^1 - r_2^2)
   \rbrack \right\}.
\label{9.2}
\end{eqnarray}
We can further rewrite this formula in terms of the creation and annihilation
operators $a^{\dagger}$ and $a$:
\begin{eqnarray}
   \lefteqn{ H^{\matrix}_{\rm XXZ} = \frac{1}{2N\sin\gamma} \left\{ 2 \left( a^{\dagger\mu_2}_{\mu_1}(1)
   a^{\dagger\nu_2}_{\mu_2}(2) a^{\nu_1}_{\nu_2}(1) a^{\mu_1}_{\nu_1}(2) \right. \right. } \nonumber \\
   & & \left. + a^{\dagger\mu_2}_{\mu_1}(2) a^{\dagger\nu_2}_{\mu_2}(1) a^{\nu_1}_{\nu_2}(2) a^{\mu_1}_{\nu_1}(1) 
   \right) \nonumber \\
   & & + \cos \gamma \left( a^{\dagger\mu_2}_{\mu_1}(1) a^{\dagger\nu_2}_{\mu_2}(1) 		
   a^{\nu_1}_{\nu_2}(1) a^{\mu_1}_{\nu_1}(1) - a^{\dagger\mu_2}_{\mu_1}(1)
   a^{\dagger\nu_2}_{\mu_2}(2) a^{\nu_1}_{\nu_2}(2) a^{\mu_1}_{\nu_1}(1) \right.
   \nonumber \\
   & & - \left. a^{\dagger\mu_2}_{\mu_1}(2) a^{\dagger\nu_2}_{\mu_2}(1) 
   a^{\nu_1}_{\nu_2}(1) a^{\mu_1}_{\nu_1}(2) + a^{\dagger\mu_2}_{\mu_1}(2)
   a^{\dagger\nu_2}_{\mu_2}(2) a^{\nu_1}_{\nu_2}(2) a^{\mu_1}_{\nu_1}(2) \right)
   \nonumber \\
   & & + i \sin\gamma \left[ \coth\xi_- ( 
   \bar{q}^{\dagger\mu_1} a^{\dagger\mu_2}_{\mu_1}(1) a^{\mu_3}_{\mu_2}(1) \bar{q}_{\mu_3}
   - \bar{q}^{\dagger\mu_1} a^{\dagger\mu_2}_{\mu_1}(2) a^{\mu_3}_{\mu_2}(2) \bar{q}_{\mu_3}   
   ) \right. \nonumber \\
   & & \left. \left. \coth\xi_+ (
   q^{\dagger}_{\mu_1} a^{\dagger\mu_1}_{\mu_2}(1) a^{\mu_2}_{\mu_3}(1) q^{\mu_3}
   - q^{\dagger}_{\mu_1} a^{\dagger\mu_1}_{\mu_2}(2) a^{\mu_2}_{\mu_3}(2) q^{\mu_3}
   \right] \right\}.
\label{9.3}
\end{eqnarray}
\item {\em integrable spin-$\frac{1}{2}$ XYZ model} \cite{InKo}.  The Hamiltonian is
\begin{eqnarray}
   \lefteqn{ H^{\spin}_{\rm XYZ} = \sum_{j=1}^{c-1} \left\{ \left[ (1 +
\Gamma)
   \tau_j^x \tau_{j+1}^x + (1 - \Gamma) \tau_j^y \tau_{j+1}^y + 
   \Delta \tau_j^z \tau_{j+1}^z \right] \right. } \nonumber \\
   & & + {\rm sn}\eta \left( A_- \tau_1^z + B_- \tau_1^+ + C_- \tau_1^- +
   A_+ \tau_c^z + B_+ \tau_c^+ \right. \nonumber \\
   & & \left. \left. C_+ \tau_c^- \right) \right\},
\end{eqnarray}
where
\begin{eqnarray*}
   & & \tau^{\pm} = \tau^x \pm i \tau^y \\
   & & \Gamma = k {\rm sn}^2 \eta, \; \Delta = {\rm cn} \eta {\rm dn} \eta,
\\
   & & A_{\pm} = \frac{{\rm cn} \xi_{\pm} {\rm dn} \xi_{\pm}}{{\rm sn}
\xi_{\pm}},  \;
   B_{\pm} = \frac{2 \mu_{\pm} (\lambda_{\pm} + 1)}{{\rm sn} \eta_{\pm}}, \;
   C_{\pm} = \frac{2 \mu_{\pm} (\lambda_{\pm} - 1)}{{\rm sn} \eta_{\pm}},
\end{eqnarray*}
and ${\rm sn} u = {\rm sn} (u; k)$, the Jacobi elliptic function of modulus $0
\leq k \leq 
1$.  $\eta$, $\xi_{\pm}$ and $\lambda_{\pm}$ are arbitrary constants.  The
corresponding 
Hamiltonian in the matrix model is
\begin{eqnarray}
   \lefteqn{ H^{\matrix}_{\rm XYZ} = 2 (\s_{12}^{21} + \s_{21}^{12}) + 
   2 \Gamma (\s_{11}^{22} + \s_{22}^{11}) } \nonumber \\
   & & + \Delta (\s_{11}^{11} - \s_{12}^{12} - \s_{21}^{21} + \s_{22}^{22}) 
   + {\rm sn} \eta \left[ A_- (l_1^1 - l_2^2) + B_- l_2^1 + C_- l_1^2 \right. \nonumber \\
   & & \left. + A_+ (r_1^1 - r_2^2) + B_+ r_2^1 + C_+ r_2^1 \right].
\end{eqnarray}
\item {\em integrable spin-1 XXZ model, or Fateev-Zamolodchikov model}
\cite{ZaFa}.  The 
Hamiltonian is
\begin{eqnarray}
   \lefteqn{H^{\spin}_{\rm FZ} = \sum_{j=1}^{c-1} \left\{ {\bf S}_j \cdot {\bf
S}_{j+1} -
   ( {\bf S}_j \cdot {\bf S}_{j+1} )^2 + \frac{1}{2} (q - q^{-1})^2 S_j^z
S_{j+1}^z 
   \right. } \nonumber \\
   & & - \frac{1}{2} (q - q^{-1})^2 (S_j^z S_{j+1}^z)^2 - (q + q^{-1} - 2)
   \left[ (S_j^x S_{j+1}^x + S_j^y S_{j+1}^y) S_j^z S_{j+1}^z \right. \nonumber
\\
   & & \left. + \left. S_j^z S_{j+1}^z (S_j^x S_{j+1}^x + S_j^y S_{j+1}^y)
\right] +
   \frac{1}{2} (q - q^{-1})^2 \left[ (S_j^z)^2 + (S_{j+1}^z)^2 \right]
\right\}
   \nonumber \\
   & & + \frac{1}{2} (q^2 - q^{-2}) (S_c^z - S_1^z)
\end{eqnarray}
where $q$ is an arbitrary constant, and $S^x$, $S^y$ and $S^z$ are spin-1
matrices:
\begin{eqnarray*}
   & & S^x = \frac{1}{\sqrt{2}} \left( 
   \begin{array}{ccc} 0 & 1 & 0 \\ 1 & 0 & 1 \\ 0 & 1 & 0 \end{array}
\right),
   S^y = \frac{1}{\sqrt{2}} \left( 
   \begin{array}{ccc} 0 & -i & 0 \\ i & 0 & -i \\ 0 & i & 0 \end{array}
\right)
   \mbox{, and} \\
   & & S^z = \left( \begin{array}{ccc} 1 & 0 & 0 \\ 0 & 0 & 0 \\ 0 & 0 & -1
\end{array} 
   \right).
\end{eqnarray*}
We identify
\[ \left( \begin{array}{c} 1 \\ 0 \\ 0 \end{array} \right), \;
   \left( \begin{array}{c} 0 \\ 1 \\ 0 \end{array} \right) \; \mbox{and} \;
   \left( \begin{array}{c} 0 \\ 0 \\ 1 \end{array} \right) \]
in the spin-1 chain model by the numbers 1, 2 and 3 respectively in the matrix
model.  Then 
the corresponding Hamiltonian in the matrix model is
\begin{eqnarray}
   \lefteqn{ H^{\matrix}_{\rm FZ} = ( - \s_{12}^{12} + \s_{12}^{21} + \s_{21}^{12} - \s_{21}^{21} ) 
   + ( - \s_{23}^{23} + \s_{23}^{32} + \s_{32}^{23} - \s_{32}^{32} ) }
   \nonumber \\
   & & - 2 (\s_{13}^{13} + \s_{31}^{31}) - (\s_{13}^{31} + \s_{31}^{13} + 2 \s_{22}^{22}) \nonumber \\
   & & - \frac{1}{2} (q^2 + q^{-2}) (\s_{13}^{13} + \s_{31}^{31}) +
   (q + q^{-1}) (\s_{13}^{22} + \s_{31}^{22} + \s_{22}^{13} + \s_{22}^{31}) \nonumber \\
   & & + \frac{1}{2} (q - q^{-1})^2 (2 \s_1^1 + 2\s_3^3 - l_1^1 - r_1^1 - l_3^3 - r_3^3)
   \nonumber \\
   & & + (q^2 - q^{-2}) (r_1^1 - r_3^3 + l_1^1 - l_3^3).
\end{eqnarray}      
\item {\em integrable spin-1 XYZ model}.  The Hamiltonian was worked out by Mezincescu and 
Nepomechie \cite{MeNe}.  We can use the same trick as above to work out the corresponding 
integrable matrix model in the large-$N$ limit.
\end{itemize}

\section{Quantum Chromodynamics in Two Dimensions}
\label{s10}

The second family of models in which our symmetry algebras are applicable are QCD models in various dimensions,
dimensionally reduced or not. 
We are going to explain briefly how to obtain the approximate Hamiltonian of a (2+1)-dimensional QCD model with 1 
flavor of quarks and antiquarks by dimensional reduction from (2+1) to (1+1), which is actually a model of mesons 
studied by Antonuccio and Dalley \cite{AnDa}, and to write down the Hamiltonian as an element of the grand algebra. 
We will give explicit formulae for the Hamiltonian \footnote{In fact our equation Eq.(\ref{9.8}) do not agree with 
Eq.(14) for the self-energy in Ref.\cite{AnDa}. But since these
self-energy terms will anyway be absorbed into a redefinition of the mass of the scalar, they
don't affect the answer.}. We will then make a different approximation from ref.\cite{AnDa} 
(not assuming that the gluon number is  conserved) which will give
us a partial analytical understanding of the spectrum. 

In this section we will let the regulator \m{\Lambda\to \infty}, so
that the momentum indices will take an infinite number of values. In
order that the previous discussions of our algebra apply directly here, we will
need to regularize the field theory such that the momentum variables
can take only a finite number \m{\Lambda} of distinct values. But this
is mostly a  technicality, since  we will only talk about  field theories without
divergences for which the limit \m{\Lambda\to \infty} should exist.

Consider an SU($N$) gauge theory in (2+1) dimensions with one flavor of quarks of mass $m$.  Let $\tilde{g}$ be 
the strong coupling constant, $\alpha$ and $\beta$ 
be ordinary spacetime indices ($\alpha$ and $\beta\in\{ 0, 1, 3 \}$), $A_{\alpha}$ be a 
gauge potential and $\Psi$ be a quark field in the fundamental representation of the gauge 
group U($N$).  $A_{\alpha}$ is a traceless $N \times N$ Hermitian matrix fields whereas 
$\Psi$ is a column vector of $N$ Grassman fields.  Let the covariant derivative
\m{ D_{\alpha}\Psi = \partial_{\alpha} \Psi + {\rm i} A_{\alpha} \Psi }
and the Yang-Mills field be 
\m{ F_{\alpha\beta} = \partial_{\alpha} A_{\beta} - \partial_{\beta} A_{\alpha}
   + i \lbrack A_{\alpha}, A_{\beta} \rbrack } as usual.
Then the action is
\begin{equation}
   S = \int d^4 x \left[ - \frac{1}{4\tilde{g}^2} {\rm Tr} F_{\alpha\beta} F^{\alpha\beta}
   + {\rm i} \bar{\Psi} \gamma^{\alpha} D_{\alpha} \Psi - m \bar{\Psi} \Psi \right]
\end{equation}
in the Weyl representation
\begin{equation}
   \gamma^0 = \left( \begin{array}{cc} 0 & - {\bf 1} \\ - {\bf 1} & 0 \end{array} \right)
   \; \mbox{and} \;
   \gamma^i = \left( \begin{array}{cc} 0 & \tau^i \\ - \tau^i & 0 \end{array} \right)
\end{equation}
for $i$ = 1 or 3.  Following the same procedure in Ref.\cite{AnDa}, we can get the following expressions for the
light-front momentum $P^-$ and energy $P^+$:
\begin{eqnarray}
   \lefteqn{ P^+ = \int_0^{\infty} dk \, k \left[ 1 \otimes \s^k_k \otimes 1 + \sum_{j = +, -} ( 
   \bar{\Xi}^{kj}_{kj} \otimes l^{\phi}_{\phi} \otimes 1 \right. } \nonumber \\
   & & \left. + 1 \otimes r^{\phi}_{\phi} \otimes \Xi^{kj}_{kj} ) \right]
\label{9.7} \\
   \lefteqn{ P^- = \int_0^{\infty} dk \, h_{IV}(k) 1 \otimes \s^k_k \otimes 1 + 
   \int_0^{\infty} dk_1 dk_2 dk_3 dk_4 \cdot } \nonumber \\
   & & \left[ h_{IV}(k_1, k_2; k_3, k_4) \delta(k_1 + k_2 - k_3 - k_4) 1 \otimes \s^{k_3 k_4}_{k_1 k_2} \otimes 1 
   \right. \nonumber \\
   & & + h_{IV}(k_1; k_2, k_3, k_4) \delta(k_1 - k_2 - k_3 - k_4) \left( 
   1 \otimes \s^{k_2 k_3 k_4}_{k_1} \otimes 1 \right. \nonumber \\
   & & \left. \left. + 1 \otimes \s^{k_1}_{k_2 k_3 k_4} \otimes 1 \right) \right]
   \nonumber \\
   & & + \sum_{j=+,-} \int_0^{\infty} dk h_{II}(k) \left( \bar{\Xi}^{kj}_{kj} \otimes l^{\emptyset}_{\emptyset} 
   \otimes 1 + 1 \otimes r^{\emptyset}_{\emptyset} \otimes \Xi^{kj}_{kj} \right) \nonumber \\
   & & + \int_0^{\infty} dk_1 dk_2 dk_3 h_{II}(k_1; k_2, k_3) \delta(k_1 - k_2 - k_3)
   \left( - \bar{\Xi}^{k_1,+}_{k_3,-} \otimes l^{\emptyset}_{k_2} \otimes 1 \right. \nonumber \\
   & & + \bar{\Xi}^{k_1,-}_{k_3,+} \otimes l^{\emptyset}_{k_2} \otimes 1
   - \bar{\Xi}^{k_3,-}_{k_1,+} \otimes l^{k_2}_{\emptyset} \otimes 1 
   + \bar{\Xi}^{k_3,+}_{k_1,-} \otimes l^{k_2}_{\emptyset} \otimes 1  \nonumber \\
   & & - 1 \otimes r^{\emptyset}_{k_2} \otimes \Xi_{k_3,+}^{k_1,-} 
   + 1 \otimes r^{\emptyset}_{k_2} \otimes \Xi_{k_3,-}^{k_1,+} 
   - 1 \otimes r^{k_2}_{\emptyset} \otimes \Xi_{k_1,-}^{k_3,+}
   \nonumber \\
   & & \left. + 1 \otimes r^{k_2}_{\emptyset} \otimes \Xi_{k_1,+}^{k_3,-} \right) \nonumber \\
   & & + \sum_{j=+,-} \int_0^{\infty} dk_1 dk_2 dk_3 dk_4 \left[ h_{II}(k_1, k_2; k_3, k_4) 
   \delta(k_1 + k_2 - k_3 - k_4) \cdot \right. \nonumber \\
   & & \left( \bar{\Xi}^{k_1,j}_{k_4,j} \otimes l^{k_2}_{k_3} \otimes 1 +
   1 \otimes r^{k_2}_{k_3} \otimes \Xi_{k_4, j}^{k_1, j} \right) \nonumber \\
   & & + h_{II}(k_1; k_2, k_3, k_4) \delta(k_1 - k_2 - k_3 - k_4) \left(
   \bar{\Xi}^{k_1,j}_{k_4,j} \otimes l^{\emptyset}_{k_3 k_2} \otimes 1 + \right.\nonumber\\
   & & \left. \left. + \bar{\Xi}^{k_4,j}_{k_1,j} \otimes l^{k_3 k_2}_{\emptyset} \otimes 1 +
   1 \otimes r^{\emptyset}_{k_2 k_3} \otimes \Xi_{k_4, j}^{k_1, j} +
   1 \otimes r^{k_2 k_3}_{\emptyset} \otimes \Xi_{k_1, j}^{k_4, j} \right)  \right] \nonumber \\
   & & + \int_0^{\infty} dk_1 dk_2 dk_3 dk_4 h_I(k_1, k_2; k_3, k_4) \delta(k_1 + k_2 - k_3 - k_4) \cdot 
   \nonumber \\ 
   & & \left( \bar{\Xi}^{k_1, +}_{k_3, +} \otimes f^{\emptyset}_{\emptyset} \otimes \Xi_{k_4, -}^{k_2, -}
   + \bar{\Xi}^{k_1, -}_{k_3, -} \otimes f^{\emptyset}_{\emptyset} \otimes \Xi_{k_4, +}^{k_2, +} \right) 
\label{9.8}
\end{eqnarray}
where
\begin{eqnarray*}
   h_{IV}(k) & = & \frac{g^2 N}{4 \pi k} \int_0^k dp \frac{(k+p)^2}{p(k-p)^2}; \\
   h_{IV}(k_1, k_2; k_3, k_4) & = & \frac{g^2 N}{8 \pi} \frac{1}{\sqrt{k_1 k_2 k_3 k_4}} \cdot \\
   & & \left[ \frac{(k_2 - k_1) (k_4 - k_3)}{(k_2 + k_1)^2}
   - \frac{(k_3 + k_1) (k_4 + k_2)}{(k_4 - k_2)^2} \right]; \\
   h_{IV}(k_1; k_2, k_3, k_4) & = & \frac{g^2 N}{8 \pi} \frac{1}{\sqrt{k_1 k_2 k_3 k_4}} \cdot \\
   & & \left[ \frac{(k_2 + k_1) (k_4 - k_3)}{(k_4 + k_3)^2}
   - \frac{(k_4 + k_1) (k_3 - k_2)}{(k_3 + k_2)^2} \right]; \\
   h_{II}(k) & = & \frac{m^2}{2k} + \frac{g^2 N}{4 \pi} \int_0^k dp \left[ \frac{1}{p^2} + 
   \frac{1}{2 p (k - p)} + \frac{1}{(k-p)^2} \right]; \\
   h_{II}(k_1; k_2, k_3) & = & \frac{g}{4} \sqrt{\frac{N}{\pi}} \frac{m}{\sqrt{k_2}} 
   (\frac{1}{k_1} - \frac{1}{k_3}); \\
   h_{II}(k_1, k_2; k_3, k_4) & = & \frac{g^2 N}{8 \pi} \frac{1}{\sqrt{k_2 k_3}}
   \left[ - \frac{2(k_2 + k_3)}{(k_2 - k_3)^2} + \frac{1}{k_1 + k_2} \right]; \\
   h_{II}(k_1; k_2, k_3, k_4) & = & \frac{g^2 N}{8 \pi} \frac{1}{\sqrt{k_2 k_3}}
   \left[ \frac{2(k_2 - k_3)}{(k_2 + k_3)^2} + \frac{1}{k_3 + k_4} \right] \mbox{; and} \\
   h_I(k_1, k_2; k_3, k_4) & = & - \frac{g^2 N}{2 \pi} \frac{1}{(k_1 - k_3)^2}.
\end{eqnarray*}
(remark: $h_{IV}(k)$, $h_{IV}(k_1, k_2; k_3, k_4)$ and $h_{IV}(k_1; k_2, k_3, k_4)$ are closely
related to the coefficients $A$, $B$ and $C$ in \cite{DaKl93a}.)
The Roman numerals $I$ and $IV$ carried by some $h$'s refer to the fact that these are coefficients of operators of
the first and fourth kinds respectively, whereas the Roman number $II$ carried by other $h$'s signify that these
are coefficients of operators of the second and third kinds.
(Eq.(\ref{9.8}) corresponds to Eq.(14) in Ref.\cite{AnDa}, which shows terms corresponding to operators of the 
second kind only.)

In the limit of a large number of gluons, we can regard each gluon
as moving in the mean field created by the others. Hence it should be
a sensible approximation to consider them all as occupying the same
state with wave function \m{u(k)}. In this case we effectively have a
one-dimensional Hilbert space for the gluons, and our algebras
simplify to the case \m{\Lambda=1}. 

Suppose we choose an orthonormal basis where the first element is \m{u(k)}. 
Then we can split our Hamiltonian as 
\beq
	P^-= P_0^-+P_1^-
\eeq
where \m{P_0^-} is the contribution when all the indices correspond to \m{u(k)} and \m{P_1^-} is the rest of the 
terms. In fact,
\beq
P_0^-=\mu \sigma_1^1+\alpha\sigma^2_2+\beta\sigma^1_3+\beta^{\ast}\sigma^3_1.
\eeq
The last two terms in \m{P_0^-} describe the processes where a gluon decays into three
 gluons or three of them combine into one gluon.
The coefficients are integrals over the wavefunction \m{u(k)}:
\beq
   \mu & = & \int_0^{\infty} dk h_{IV}(k) u^{\ast}(k) u(k); \\
   \alpha & = & \int_0^{\infty} dk_1 dk_2 dk_3 dk_4 h_{IV}(k_1, k_2; k_3, k_4) u^{\ast}(k_3) u^{\ast}(k_4) 
   u(k_1) u(k_2);
\eeq
and
\beq
   \beta = \int_{\infty} dk_1 dk_2 dk_3 dk_4 h_{IV}(k_1; k_2, k_3, k_4) u^{\ast}(k_1) u(k_2) u(k_3) u(k_4). 
\eeq

We can then
determine the spectrum of \m{P_0^-} for a given \m{u(k)} using the
\m{\Lambda=1} special case of our commutation relations. It amounts
to solving some recursion relations which we omit for the sake of brevity.

The function \m{u(k)} can  be determined by mean field theory. This is the same as
picking the \m{u(k)} that minimizes the expectation value
\m{<P^->}, in each sector with a given average number of gluons.  A
moments thought will show that \m{<P_1^->=0}. Hence we must find the
spectrum of \m{P_0^-} and minimize the energy over all \m{u(k)}.

Clearly the sectors with even and odd numbers of gluons do not mix with each
other. The spectrum of \m{P_0^-} in the even sector is
\beq
   \lambda_n= \mu + (1 + 2n) \sqrt{(\mu + \alpha)^2 - 4 \beta^2}, \quad n=0,1,2,\ldots
\eeq
 
The `principal quantum number' \m{n} is approximately the average number of
gluons.  The wave function is a linear combination with differing numbers of gluons:
\beq
	|\psi_n \rangle = \sum_{r=1}^\infty \xi_{nr}|2r \rangle
\eeq
where the coefficients $\xi_{nr}$ depend on $\mu$, $\alpha$ and $\beta$.  
The lowest energy state has an exponentially decreasing probability to have
several gluons, in agreement with previous studies.
But as \m{n} grows, the wave function has
contributions from states with different numbers of gluons.

The linearly rising eigenvalues of \m{P^-} correspond to a mass
spectrum that is also approximately linearly rising:
\beq
	M_n^2=\lambda_n P_n^+
\eeq
where
\beq
	P_n^+ = \int_0^{\infty} dk k u^{\ast}(k) u(k).
\eeq
We can then determine \m{u(k)} by minimizing the energy for each fixed \m{n}.
We will report more detailed information on the spectrum in later papers.

\vskip 1pc
\noindent \Large{\bf \hskip .2pc Acknowledgments}
\vskip 1pc
\noindent 

\normalsize
We thank O. T. Turgut for discussions at an early stage of this
work. S. G. R. thanks the I.H.E.S., where part of this work was done,
 for hospitality. We were  supported in part by funds provided by the U.S.
Department of Energy under grant DE-FG02-91ER40685.

\vskip 1pc
\noindent \Large{\bf \hskip .2pc Appendix}
\vskip 1pc

\normalsize
\appendix
\section{On Multi-Indices}
\label{s1-1}

Much of our work involves manipulating tensors carrying multiple indices.
For the convenience of the reader, we give here a summary of the notations used
in this paper for multi-indices.

We will use lower case Latin letters such as \m{i,j, i_1, j_2 } to
denote indices with values 1, 2, \ldots, or $\Lambda$. Here \m{\Lambda}
itself is a fixed positive integer, denoting the number of degrees of
freedom of gluons. Often we will have
to deal with a whole sequence  of indices \m{i_1i_2i_3\ldots i_a},
which we will denote by the corresponding uppercase letter \m{I} as the collective
index of this sequence. The
length of the  sequence \m{I} will be denoted by \m{\#(I)}. 

Thus if  
\[ I = i_1 i_2 \ldots i_a \]
and 
\[ J = j_1 j_2 \ldots j_b, \]
then we have  $\#(I) = a$ and $\#(J) = b$.  The composition  of these two
sequences will be denoted by 
\[ I J = i_1 i_2\ldots i_a j_1 j_2 \ldots j_b. \]
In particular, we have 
\beq
	Ij=i_1i_2\ldots i_aj,
\eeq
when only a single index is added to the end. This operation of
composing sequences is associative but not commutative:
\beq
	IJ\neq JI, \quad I(JK)=(IJ)K=IJK.
\eeq

Often we will have to allow the null sequence \m{\emptyset} among the range of
values of a collective index. A collective index that is allowed to take the null sequence
as its value will have  a dot over it. Thus the
possible values of \m{\dot{I}} are 
\beq
   \dot{I} = \emptyset,1,2,\ldots \Lambda,11,12,\ldots
             1\Lambda,21,\ldots, \Lambda\Lambda,\ldots
\eeq
while those of \m{I} do not include the empty set:
\beq
	I=1,2,\cdots \Lambda,11,12,\cdots
1\Lambda,21,\cdots, \Lambda\Lambda,\cdots
\eeq
Of course the length of the null sequence is zero. It is the identity
element of the composition law above,
\beq
	\emptyset\dot{I}=\dot{I}\emptyset=\dot{I}.
\eeq
In fact the inclusion of the empty sequence in the set of all
sequences turns it into  a semi-group under the above
composition law.

The equation $I = J$ means that they have the same length $a$ (say) and
\[ i_1 = j_1; i_2 = j_2; \ldots i_a=j_a. \] 
In the same way we can define \m{\dot{I}=\dot{J}} either if they are
both the empty sequence, or if they are equal in the above sense.

The Kronecker delta function for integer sequences is defined as follows:
\[ \delta^I_J \equiv \left\{ \begin{array}{ll}
   				1 & \mbox{if $I = J$; or} \\
   				0 & \mbox{if $I \neq J$.} \\
   			     \end{array} \right\} \]
and similarly for dotted indices.
The summation sign in an expression such as 
\[ \sum_I X^I_J, \]  means that all possible distinct sequences, {\em excluding
the empty sequence}, are summed over.  (In all practical cases, it turns out
that there is only a finite number of $I$'s such that $X^I_J \neq 0$ so there
will be no convergence problems.) 
On the other hand, the summation sign over a dotted index 
\[ \sum_{\dot{I}} X^{\dot{I}}_J \]
means that  all possible distinct sequences for $\dot{I}$, {\em including the
empty sequence}, are summed over.

Often we will have to sum over all the ways of splitting a sequence
into subsequences.  For example, the summation sign on the L.H.S. of the equation
\beq
   \sum_{I_1 I_2 = I} X^{I_1 }Y^{I_2}_K \equiv 
   \sum_{I_1I_2}\delta_I^{I_1I_2}X^{I_1}Y^{I_2}_K
\label{1-1.1}
\eeq
denotes the sum over all the ways in which a given index \m{I}
can be split into two {\em nonempty} subsequences \m{I_1} and
\m{I_2}. If there is no way to split $I$ as required, then the sum simply yields 0.
For example, Eq.(\ref{1-1.1}) yields 0 if $I$ has only one integer.
If the first subsequence is allowed to be empty in the sum, we would
write instead
\beq
   \sum_{\dot{I}_1 I_2 = I} X^{\dot{I}_1 }Y^{I_2}_K .
\eeq

When we talk of an algebra spanned by operators such as 
\m{f^{\dot{I}}_{\dot{J}}, l^{\dot{I}}_{\dot{J}}, r^{\dot{I}}_{\dot{J}}} or
\m{\s^I_J} \footnote{for definitions see the main text.}, the underlying vector space 
is that of {\em finite} linear combinations. For example, in the linear combination \m{\sum_{IJ}c^J_I\sigma^I_J}, 
although \m{I,J} can take an infinite number of values, only a finite number of
the complex numbers \m{c^J_I} can be non-zero. Thus there are never
any issues of convergence in the sums of interest to us: they are all
finite sums.

\end{document}

%% file: table1.tex
\begin{table}
\centerline{
\begin{tabular}[t]{|| c | c | l | rcl ||} \hline
operators    & extended 	        & \multicolumn{1}{c|}{comment}
& \multicolumn{3}{c||}	{quotient algebra}			   	      \\
             & algebra                  &                                       & 		      &   &                 			       \\ \hline \hline
$f^{\dot{I}}_{\dot{J}}$      & ${\mathit F}_{\Lambda}$      &${\mathit F}_{\Lambda}\equiv  gl_{+\infty}$&  & &  \\ \hline
$l^{\dot{I}}_{\dot{J}}$      & $\hat{\mathit L}_{\Lambda}$  & ${\mathit F}_{\Lambda}$ is a proper
             						    & ${\mathit L}_{\Lambda}$ & $\equiv$ & $\hat{\mathit L}_{\Lambda} / {\mathit F}_{\Lambda}$  \\
$f^{\dot{K}}_{\dot{L}} = l^{\dot{K}}_{\dot{L}} -  l^{\dot{K}j}_{\dot{L}j}$& & ideal of $\hat{\mathit L}_{\Lambda}$. 	        & 		      &	  &					  	      \\ \hline
$r^{\dot{I}}_{\dot{J}}$      & $\hat{\mathit R}_{\Lambda}$  & ${\mathit F}_{\Lambda}$ is a proper & ${\mathit R}_{\Lambda}$ & $\equiv$ & $\hat{\mathit R}_{\Lambda} / {\mathit F}_{\Lambda}$  \\
$f^{\dot{K}}_{\dot{L}} = r^{\dot{K}}_{\dot{L}} - r^{i\dot{K}}_{i\dot{L}}$ & & ideal of $\hat{\mathit R}_{\Lambda}$. 	        & 		      &	  &					  	      \\ \hline
$l^{\dot{I}}_{\dot{J}}$ and $r^{\dot{I}}_{\dot{J}}$ & $\hat{\mathit M}_{\Lambda}$ & $\hat{\mathit M} \neq \hat{\mathit L}_{\Lambda} \oplus \hat{\mathit R}_{\Lambda}$ & 
  ${\mathit M}_{\Lambda}$ & $\equiv$ & $\hat{\mathit M}_{\Lambda} / {\mathit F}_{\Lambda}$ \\ \hline \hline
\end{tabular}}
\caption{\em Relationship among $\salt$, $\hatleftix$, $\hatrightix$ and $\hatmultix$.  The summation convention is 
adopted for repeated indices in this table.}
\label{t1}
\end{table}

%% file: table1_1.tex
\begin{table}
\centerline{
\begin{tabular}[t]{|| c | c | l | rcl ||} \hline
operators    & extended 	        & \multicolumn{1}{c|}{comment}
& \multicolumn{3}{c||}	{quotient algebra(s)}			   	      \\
             & algebra                  &                                       & 		      &   &                 			       \\ \hline \hline
$f^I_J$      & ${\mathit F'}_{\Lambda}$      &${\mathit F'}_{\Lambda}\equiv  gl_{+\infty}$&  & &  \\ \hline
$l^I_J$      & $\hat{\mathit L'}_{\Lambda}$  & ${\mathit F'}_{\Lambda}$ is a
             proper ideal
                                                                                & ${\mathit L'}_{\Lambda}$ & $\equiv$ & $\hat{\mathit L'}_{\Lambda} / {\mathit F'}_{\Lambda}$  \\ 
${f}^K_L = l^K_L -  l^{Kj}_{Lj}$& & of $\hat{\mathit L'}_{\Lambda}$. 	        & 		      &	  &					  	      \\ \hline
$r^I_J$      & $\hat{\mathit R'}_{\Lambda}$  & ${\mathit F'}_{\Lambda}$ is a proper ideal & ${\mathit R'}_{\Lambda}$ & $\equiv$ & $\hat{\mathit R'}_{\Lambda} / {\mathit F'}_{\Lambda}$  \\
${f}^K_L = r^K_L - r^{iK}_{iL}$ & & of $\hat{\mathit R'}_{\Lambda}$. 	        & 		      &	  &					  	      \\ \hline
$l^I_J$ and $r^I_J$ & $\hat{\mathit M'}_{\Lambda}$ & $\hat{\mathit M'} \neq \hat{\mathit L'}_{\Lambda} \oplus \hat{\mathit R'}_{\Lambda}$ & 
  ${\mathit M'}_{\Lambda}$ & $\equiv$ & $\hat{\mathit M'}_{\Lambda} / {\mathit F'}_{\Lambda}$ \\ \hline
$\sigma^I_J$ & $\hat{\Sigma}_{\Lambda}$ & ${\mathit F'}_{\Lambda}$, $\hat{\mathit L'}_{\Lambda}$ and $\hat{\mathit R'}_{\Lambda}$ are & 
  $\Sigma_{\Lambda}$ & $\equiv$ & $\hat{\Sigma}_{\Lambda} / {\mathit F'}_{\Lambda}$
\\
  $l^I_J =\sigma^I_J-\sigma^{iI}_{iJ}$& & proper ideals of $\hat{\Sigma}_{\Lambda}$. & 
${\cal V}_{\Lambda}$ & $\equiv$ & $\hat{\Sigma}_{\Lambda} / \hat{M}_{\Lambda}'$ \\
  $r^I_J =\sigma^I_J-\sigma^{Ij}_{Jj}$& &  &
& = & $\Sigma_{\Lambda} / M_{\Lambda}'$ \\	
$f^I_J = \sigma^I_J - \sigma^{Ij}_{Jj}$ & &  & 		      &   & \\					       
$- \sigma^{iI}_{iJ} + \sigma^{iIj}_{iIj}$ & & & & & \\ \hline \hline
\end{tabular}}
\caption{\em $\hatcentrix$, its subalgebras and quotient algebras.  The summation convention is adopted
for repeated indices in this table.}
\label{t1_1}
\end{table}

%% file: table2.tex
\begin{table}
\centerline{
\begin{tabular}{|| c | c | l ||} \hline
operators & algebra & \multicolumn{1}{c||}{comment} \\ \hline \hline
$\bar{\Xi}^{\lambda_1}_{\lambda_2} \otimes f^{\dot{I}}_{\dot{J}} \otimes \Xi^{\lambda_3}_{\lambda_4}$ & 
${\mathit F}_{\Lambda, \Lambda_F}$ & 
${\mathit F}_{\Lambda, \Lambda_F} \equiv gl(\Lambda_F) \otimes {\mathit F}_{\Lambda} \otimes gl(\Lambda_F)$ \\ 
\hline
$\bar{\Xi}^{\lambda_1}_{\lambda_2} \otimes l^{\dot{I}}_{\dot{J}} \otimes 1$ & 
$gl(\Lambda_F) \otimes \hat{\mathit L}_{\Lambda}$ & \\ \hline
$1 \otimes r^{\dot{I}}_{\dot{J}} \otimes \Xi^{\lambda_3}_{\lambda_4}$ & 
$\hat{\mathit R}_{\Lambda} \otimes gl(\Lambda_F)$ & \\ \hline
$\bar{\Xi}^{\lambda_1}_{\lambda_2} \otimes f^{\dot{I}}_{\dot{J}} \otimes \Xi^{\lambda_3}_{\lambda_4}$, & 
$\hat{\mathit M}_{ \Lambda, \Lambda_F}$ & 
${\mathit F}_{\Lambda, \Lambda_F}$ is a proper ideal of $\hat{\mathit M}_{ \Lambda, \Lambda_F}$. \\
$\bar{\Xi}^{\lambda_1}_{\lambda_2} \otimes l^{\dot{I}}_{\dot{J}} \otimes 1$, and & & \\
$1 \otimes r^{\dot{I}}_{\dot{J}} \otimes \Xi^{\lambda_3}_{\lambda_4}$ & & \\ \hline
$\sigma^I_J$ & $\hat{\Sigma}_{\Lambda}$ & \\ \hline
all of the above & $\hat{\mathit G}_{\Lambda, \Lambda_F}$ & ${\mathit F}_{ \Lambda, \Lambda_F}$ and 
$\hat{\mathit M}_{ \Lambda, \Lambda_F}$ are proper ideals of \\
& & $\hat{\mathit G}_{ \Lambda, \Lambda_F}$. \\ \hline \hline
\end{tabular}}   
\caption{\em The grand algebra, its subalgebras and its ideals.}
\label{t2}
\end{table}